\documentclass[%
 reprint,
unsortedaddress,
 amsmath,amssymb,
 aps,
]{revtex4-2}

\usepackage{graphicx}
\usepackage{dcolumn}
\usepackage{bm}

\begin{document}


\title{Unravelling Heterogeneous Transport of Endosomes}

\author{Nickolay Korabel}
\affiliation{Department of Mathematics, The University of Manchester, M13 9PL, UK}
\email{nickolay.korabel@manchester.co.uk}

\author{Daniel Han}
\affiliation{
Department of Mathematics, The University of Manchester, M13 9PL, UK}%
\affiliation{
School of Biological Sciences, The University of Manchester, M13 9PT, UK}%
\affiliation{
Biological Physics, Department of Physics and Astronomy, The University of Manchester, M13 9PL, UK}%

\author{Alessandro Taloni}
\affiliation{
CNR - Consiglio Nazionale delle Ricerche, Istituto dei Sistemi Complessi, via dei Taurini 19, 00185 Roma, Italy
}%

\author{Gianni Pagnini}
\affiliation{BCAM - Basque Center for Applied Mathematics, Mazarredo 14, E-48009 Bilbao, Basque Country - Spain
}%
\affiliation{Ikerbasque - Basque Foundation for Science, Plaza Euskadi 5, E-48009 Bilbao, Basque Country - Spain
}%

\author{Sergei Fedotov}
 \affiliation{Department of Mathematics, The University of Manchester, M13 9PL, UK}

\author{Viki Allan}
\affiliation{%
School of Biological Sciences, The University of Manchester, M13 9PT, UK
}%

\author{Thomas Andrew Waigh}
\affiliation{%
Biological Physics, Department of Physics and Astronomy, The University of Manchester, M13 9PL, UK}%
\email{t.a.waigh@manchester.ac.uk}

\date{\today}

\begin{abstract}
A major open problem in biophysics is to understand the highly heterogeneous transport of many structures inside living cells, such as endosomes. We find that mathematically it is described by spatio-temporal heterogeneous fractional Brownian motion (hFBM) which is defined as FBM with a randomly switching anomalous exponent and random generalized diffusion coefficient. Using a comprehensive local analysis of a large ensemble of experimental endosome trajectories ($>{10}^5$), we show that their motion is characterized by power-law probability distributions of displacements and displacement increments, exponential probability distributions of local anomalous exponents and power-law probability distributions of local generalized diffusion coefficients of endosomes which are crucial ingredients of spatio-temporal hFBM. The increased sensitivity of deep learning neural networks for FBM characterisation corroborates the development of this multi-fractal analysis. Our findings are an important step in understanding endosome transport. We also provide a powerful tool for studying other heterogeneous cellular processes.
\end{abstract}

\maketitle


\section{\label{Intro}Introduction}

In eukaryotic cells, endosomes play a major role in sorting and transporting proteins and lipids that are taken in from the cell surface and need to be delivered to lysosomes for degradation. They also sort and redistribute material back to the cell surface or to the Golgi apparatus. Viruses and gene delivery vehicles exploit endosomal transport during infection pathways and transfection respectively \cite{Seisenberger}. Therefore, to design new antiviral drugs and develop efficient gene therapies, it is imperative to study the mechanisms of endosomal transport. The movement of endosomes along microtubules is powered by attachment to cytoplasmic dynein and various kinesin motor proteins \cite{Nielsen,Driskell}. Individual endosomes travel long intracellular distances ($\sim 1-10$  $\mu$m) in short bursts of directed motility, interspersed with periods of diffusive \cite{Rodriguez,Zajac} and sub-diffusive \cite{ELife} motion. It is likely that the heterogeneous character of endosome movement is influenced by cargo sorting and membrane fission \cite{Driskell,Rodriguez,Zajac,Rink}.

In previous works, the movement of endosomes inside living cells has been shown to follow anomalous diffusion \cite{Seisenberger,ELife,Gaard, ChenWangGranick,Kulkarni}, which is defined by the non-linear growth of the mean squared displacement (MSD), $\left< r^2 \right> \sim  t^{\alpha}$ ($\alpha \ne 1$) \cite{MetzlerKlafter}. The anomalous exponent, $\alpha$, characterizes the nature of the anomalous diffusion as slower (sub-diffusion, $\alpha < 1$) or faster (super-diffusion, $\alpha > 1$) than Brownian motion (diffusive, $\alpha = 1$). The sub-diffusion exhibited in endosomal movement can be caused by different mechanisms including: the cytoplasmic viscoelasticity produced by microtubules and organelles (e.g., the endoplasmic reticulum network); caging and crowding effects; and temporal binding. In contrast, the super-diffusive motions of endosomes are driven by molecular motors. Various anomalous diffusion models have been proposed to address different mechanisms of biological motility  \cite{MetzlerKlafter,KlagesBOOK,HF,Meroz,BGM,WaighBOOK,KlafterSokolov2011,Sokolov2012,BressloffNewby,Metzler2014,Kervrann}. Fractional Brownian motion (FBM), continuous time random walk (CTRW) and fractional Langevin equation (FLE) are among the most popular models. However, many previous models only use a constant generalized diffusion coefficient and a constant anomalous exponent to describe endosomal movement. This conflicts with the biological evidence of spatio-temporal heterogeneity in endosomal dynamics generated by the numerous intracellular processes that give rise to sub-diffusion and super-diffusion \cite{Driskell,Rodriguez,Zajac,ELife,Rink}.

Recently, it has become clear that anomalous diffusion with a constant anomalous exponent and a constant generalized diffusion coefficient fail to adequately describe intracellular transport. With improved microscopy imaging and tracking methods, the intrinsic spatial and temporal heterogeneity within individual trajectories of biological processes can be elucidated, for example: diffusion of proteins and lipids in the cell membrane \cite{Saxton}; colloidal beads diffusing along linear tubes \cite{Granick1,Granick2}; the motion of micron-sized beads in amoeba \cite{Heinrich}; labeled messenger RNA molecules in living {\it E. coli} and {\it S. cerevisiae} cells \cite{Lampo}; individual quantum dots moving in the cytoplasm of living mammalian cells \cite{Weiss2020}; Rab5 and SNX1 tagged endosomal transport \cite{ELife}; and lysosomal dynamics in living cells \cite{BaLysosomes,ELife}. Heterogeneous Brownian transport processes have only had sparse coverage in the literature including the broad distribution of diffusivities in the dynamics of the pathogen-recognition receptor dendritic cell-specific intercellular adhesion molecule 3-grabbing nonintegrin \cite{ManzoPRX}; and the exponential probability distribution of diffusivities and the Laplace probability distribution of displacements in the motion of RNA molecules in {\it E. coli} and {\it S. cerevisiae} cells \cite{Granick1,Granick2,Lampo}. These discoveries led to the development of a new class of mathematical models involving heterogeneous diffusion described as ``Brownian yet non-Gaussian diffusion". Among these models, the most popular are the so-called superstatistical Brownian motion \cite{Beck1,Beck2} and FBM \cite{Molina,Mackala}, which include a population of diffusion coefficients, position dependent diffusion coefficients \cite{Chertvy2014,Spakowitz}, and models involving diffusing diffusivities \cite{Chubynsky,Sposini,Chechkin,Metzler2020} that were recently extended to FBM \cite{Wang}. Furthermore, extending heterogeneity to the consideration of the anomalous diffusion, power-law distributions of both diffusion coefficients and displacements were found in the motion of H-NS proteins \cite{SadoonWang} and recently modelled through the joint fluctuations of both the anomalous diffusion exponents and the diffusion constants of the FBM \cite{itto-beck-2021}. A relatively sparse literature also exists on position and density dependent anomalous exponents \cite{KB2010,FK2015,FH2019} and distributed-order diffusion equations \cite{Sandev}.

In this article, we characterized anomalous transport in experimental trajectories of endosomes inside eukaryotic cells. We have used traditional statistical methods and a recently developed neural network (NN) analysis \cite{ELife} based on a deep learning feedforward neural network. Both approaches reveal heterogeneous behaviour on a single trajectory level and the NN method provides a dramatic improvement in sensitivity. Using a large ensemble of experimental trajectories, we find an exponential probability distribution of anomalous exponents, a power-law probability distribution of local generalized diffusion coefficients, and a power-law probability distribution of displacements, $X$ and displacement increments, $\Delta X$, for the endosomes. We show that endosomal movement is described by heterogeneous FBM (hFBM). In particular, the superstatistical ensemble \cite{Beck1,Beck2,Metzler2020} of FBM trajectories which accounts only for the spatial heterogeneity explains the power-law probability distributions of displacements and displacement increments of experimental endosomal trajectories. However, it does not explain the mean squared displacements of ensemble of experimental endosomes. On the contrary, the spatio-temporal heterogeneous FBM with random switching from persistent ($H>0.5$) to anti-persistent regime ($H<0.5$) (in the mathematical literature known as multifractional FBM \cite{multiFBM}) with random Hurst exponent $H$ and random generalized diffusion coefficient explains all experimental results. Our findings and methodologies shed new light on the underlying mechanism of endosomal transport in eukaryotic cells and improve our understanding of complex heterogeneous intracellular processes. 

\section{Results}

\begin{figure*}[ht]
\includegraphics[scale=0.22]{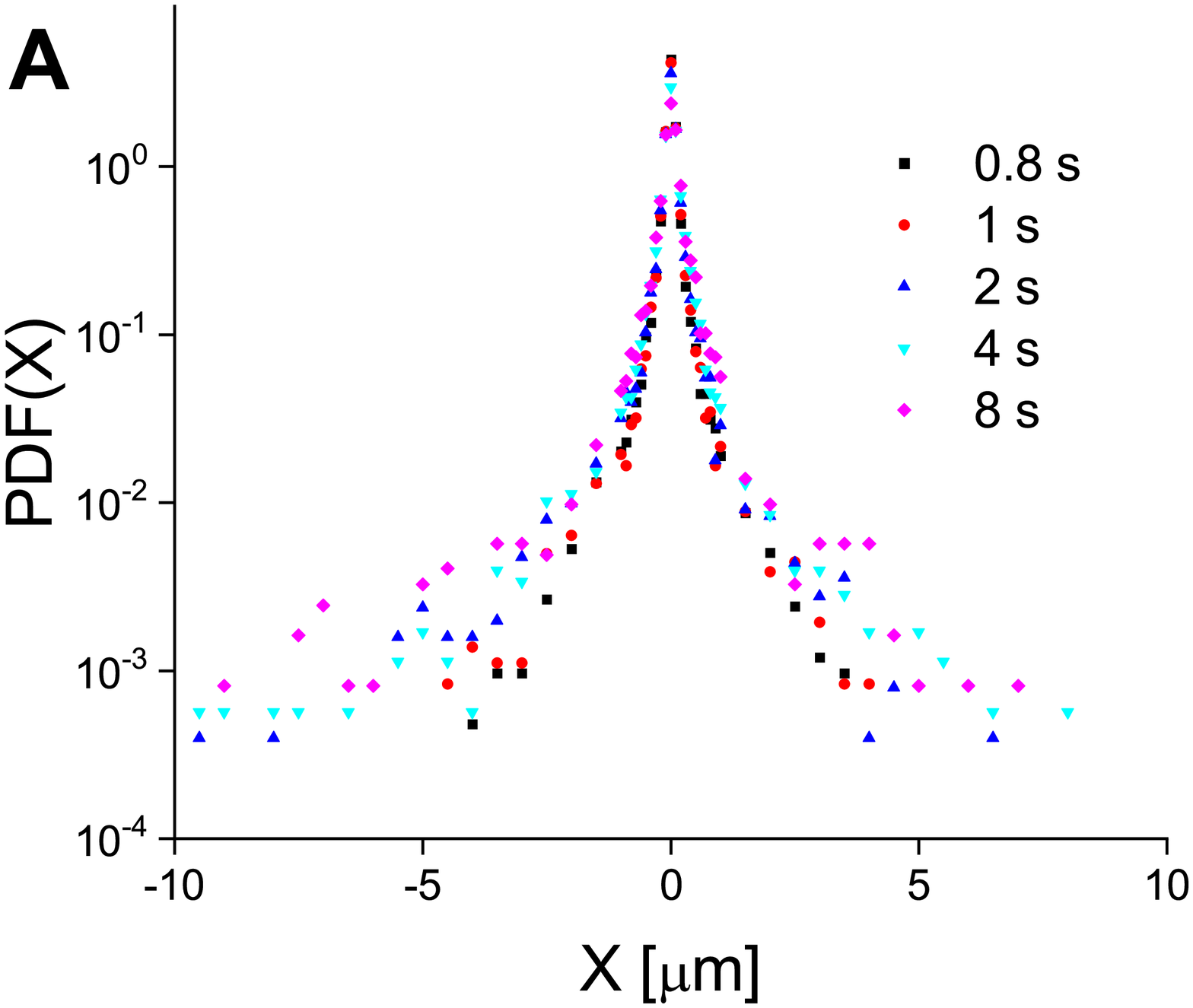}
\includegraphics[scale=0.22]{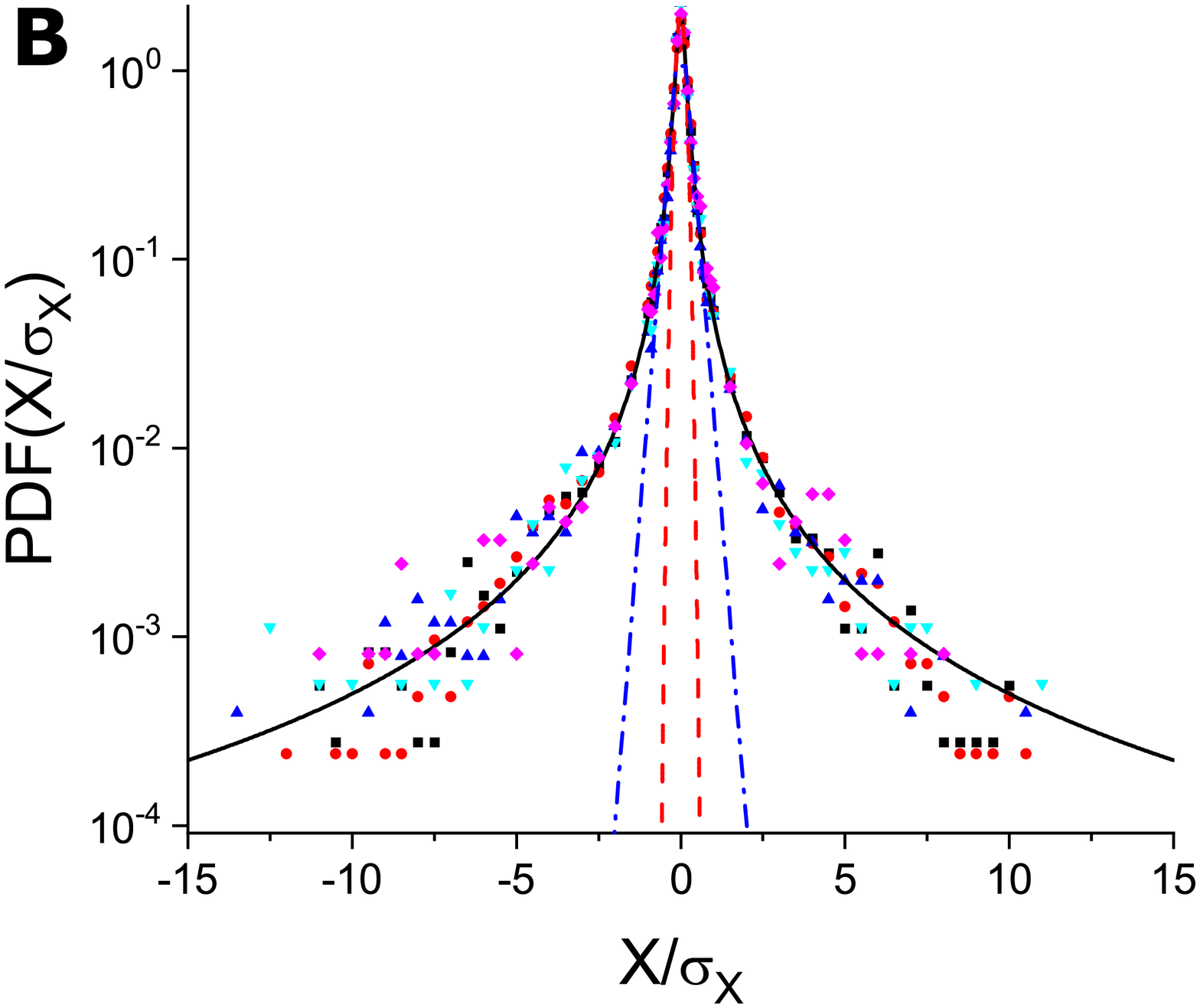}
\includegraphics[scale=0.22]{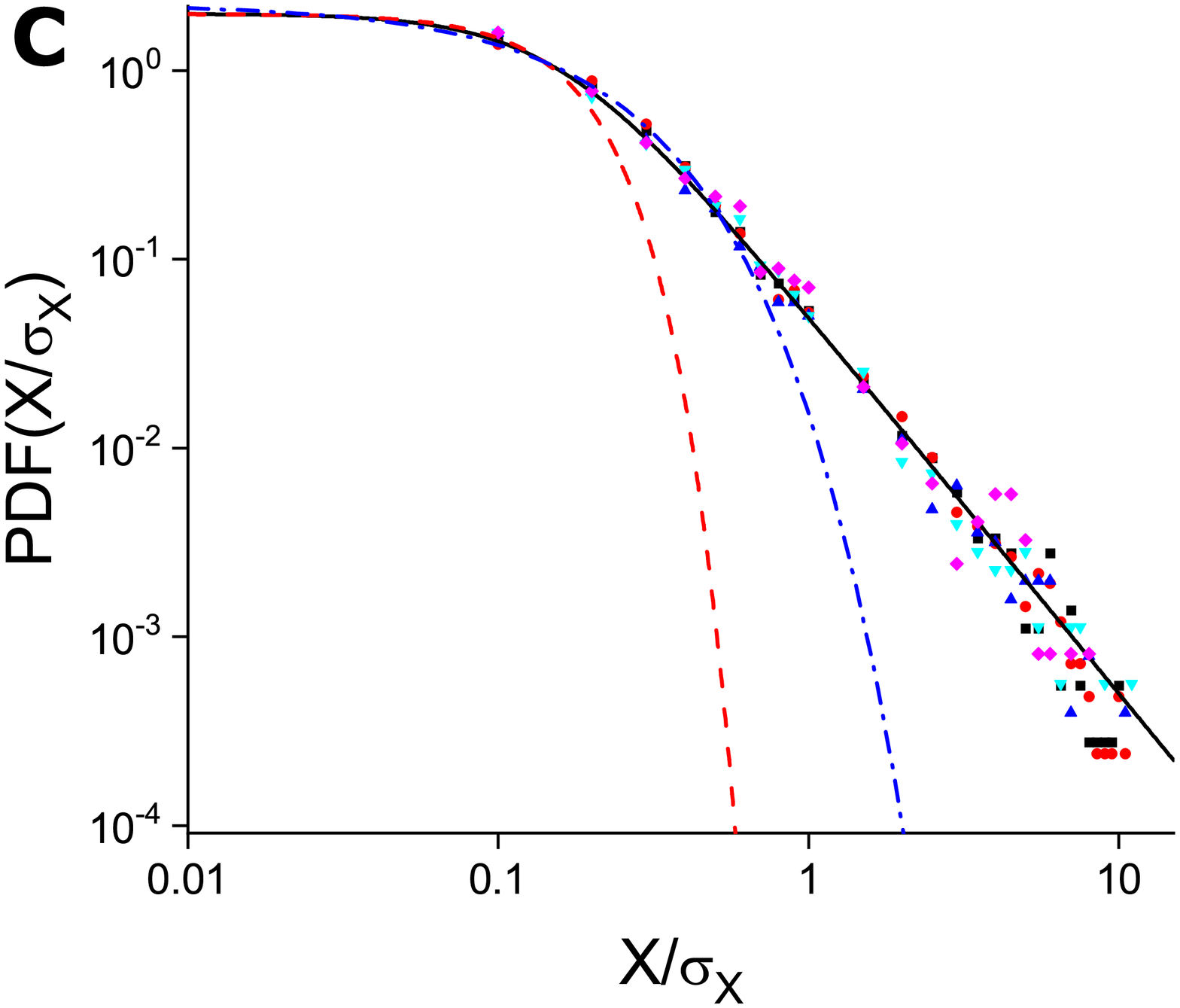}
\caption{\label{PDFX}{\bf The probability distribution of endosomal displacements is a power-law.} {\bf a.} PDFs of displacements $X=x(t)-x(0)$ extracted from experimental trajectories at $t=0.8, 1, 2, 4, 8$ seconds. {\bf b.} PDFs of displacements scaled by the standard deviation $\sigma_X$ in a log-linear scale. {\bf c.} Same as  in {\bf b} in a log-log scale. PDFs of $X/\sigma_X$ are best fitted with a power law $(X/\sigma_X)^{-2}$ (the solid curve) contrary to Gaussian (the dashed curve) and Laplace probability distributions (the dashed-dotted curve).}
\end{figure*}

\begin{figure*}[ht]
\includegraphics[scale=0.22]{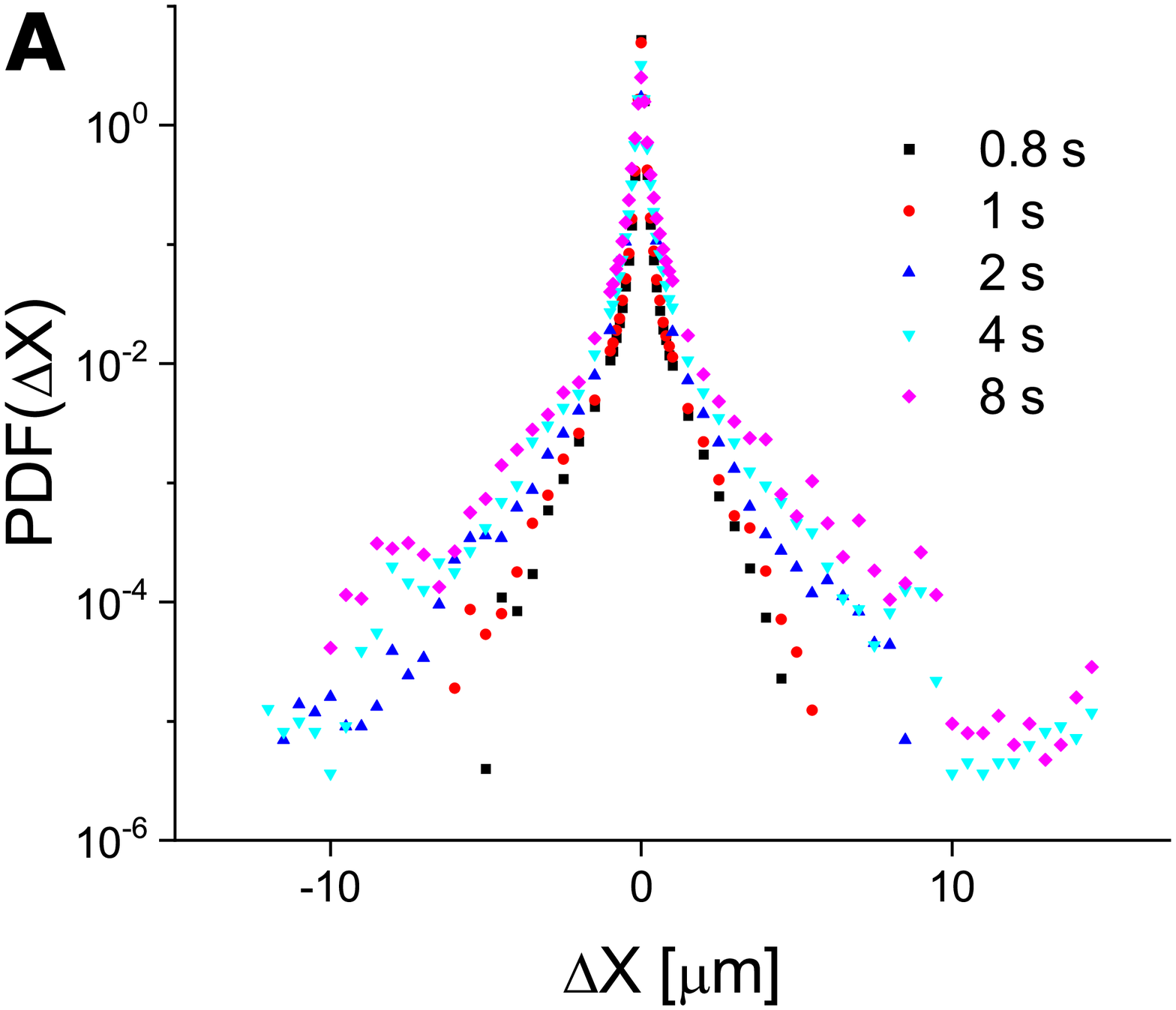}
\includegraphics[scale=0.22]{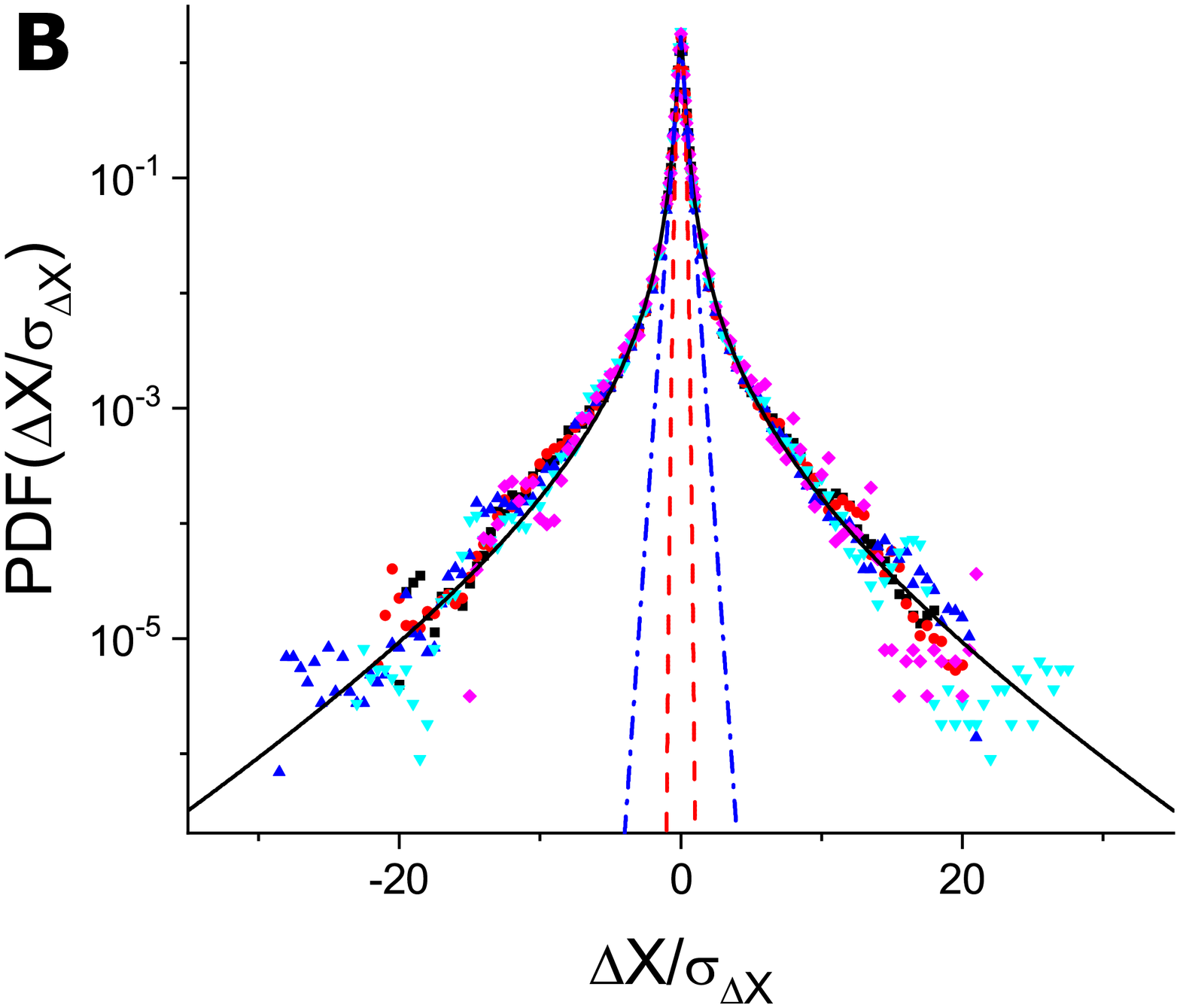}
\includegraphics[scale=0.22]{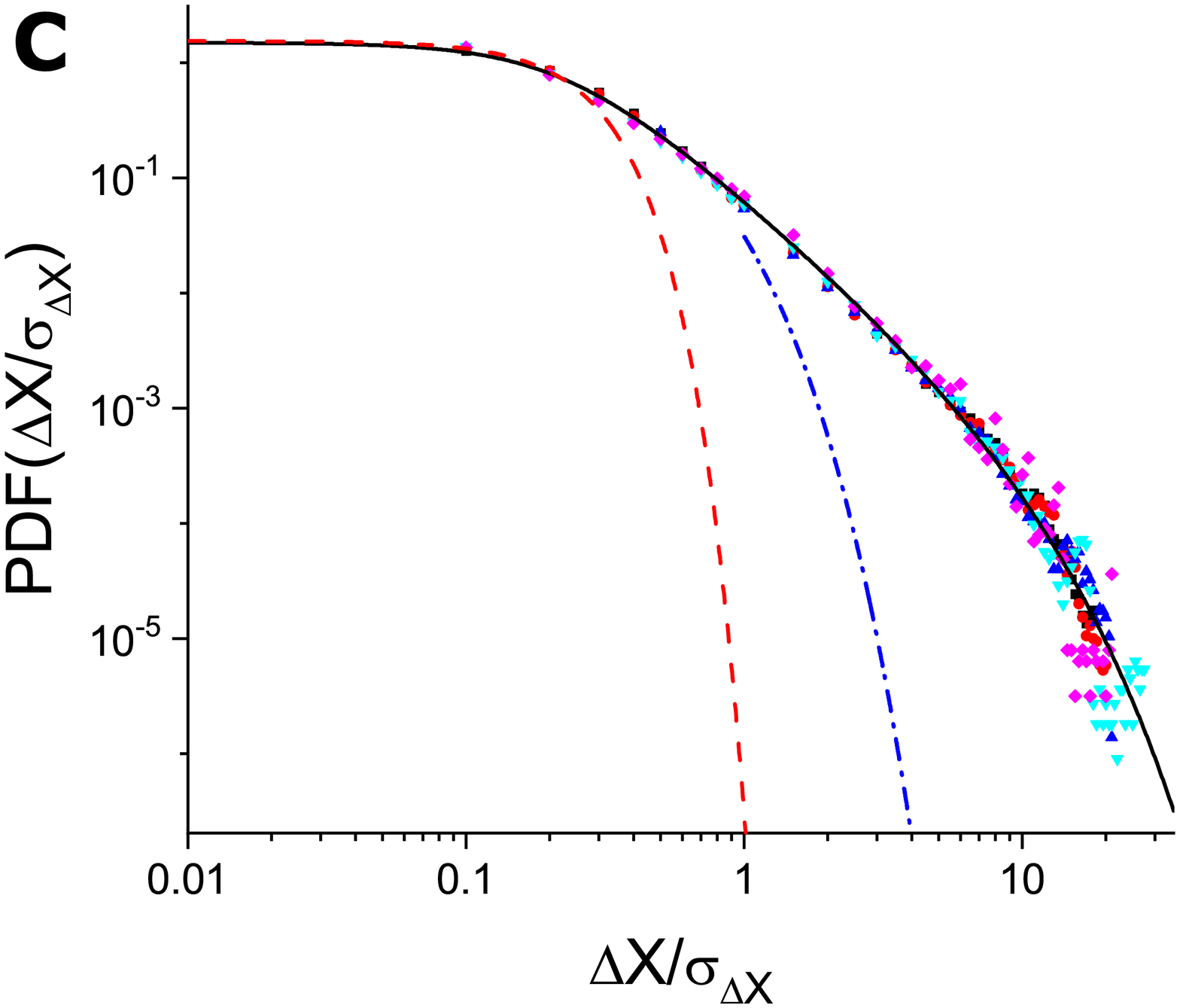}
\caption{\label{PDFDX} {\bf Probability distribution of endosomal displacement increments is a (truncated) power-law.} {\bf a.} PDFs of displacement increments $\Delta X=x(t+\Delta t)-x(t)$ for time intervals $\Delta t=0.8, 1, 2, 4, 8$ seconds. {\bf b.} Distribution of increments scaled by the standard deviation $\sigma_{dX}$ in a log-linear scale. {\bf c.} Same as in {\bf b} in a log-log scale. The solid curve represents the fit with the truncated power law, $\exp (-0.15 \Delta X/\sigma_{\Delta X}) (\Delta X/\sigma_{\Delta X})^{-2}$. The dashed curve is a fit with a Gaussian probability distribution and the dashed-dotted curve is a fit with a Laplace probability distribution. }
\end{figure*}

We studied 103,361 experimental trajectories of early endosomes in a stable MRC5 cell line expressing GFP-Rab5, obtained from tracking wide-field fluorescence microscopy videos (see \cite{ELife} for experimental details). The endosomes were tracked using an automated tracking software (AITracker, based on a convolutional neural network) \cite{Newby}. The duration of all trajectories, $T$, has a good fit to a power law distribution, $\phi\left(T\right)\sim T^{-1.85}$ (see Appendix B, Fig. S1). This distribution is a manifestation of the heterogeneity of the trajectories. Slow moving endosomes stay longer within the observation volume and therefore have longer trajectories than fast moving endosomes \cite{Etoc} leading to the emergence of the power-law probability distribution for the trajectories' duration. 

\subsection{Endosomal trajectories have power-law distributed displacements and increments}

\begin{figure*}[ht]
\includegraphics[scale=0.22]{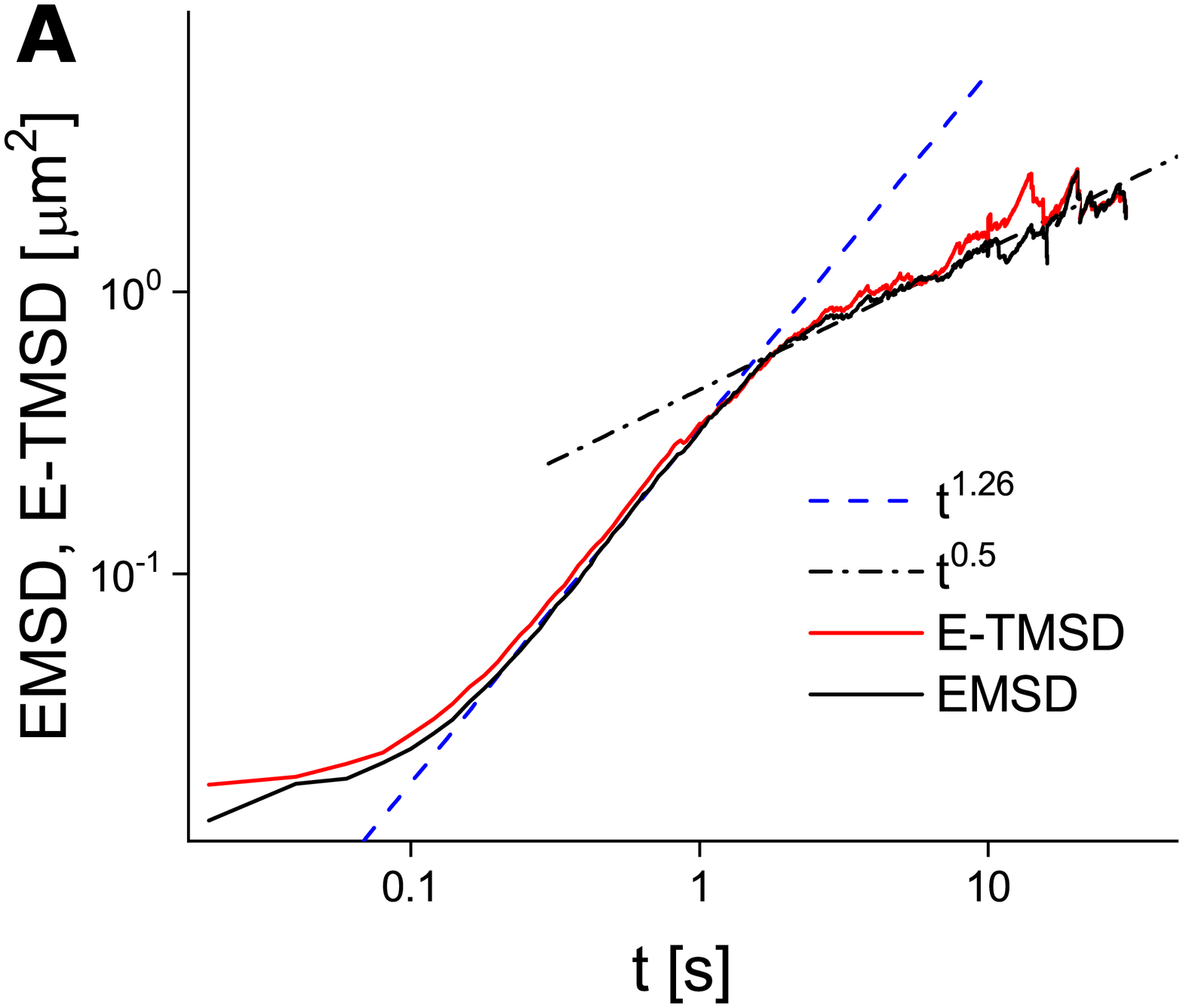}
\includegraphics[scale=0.22]{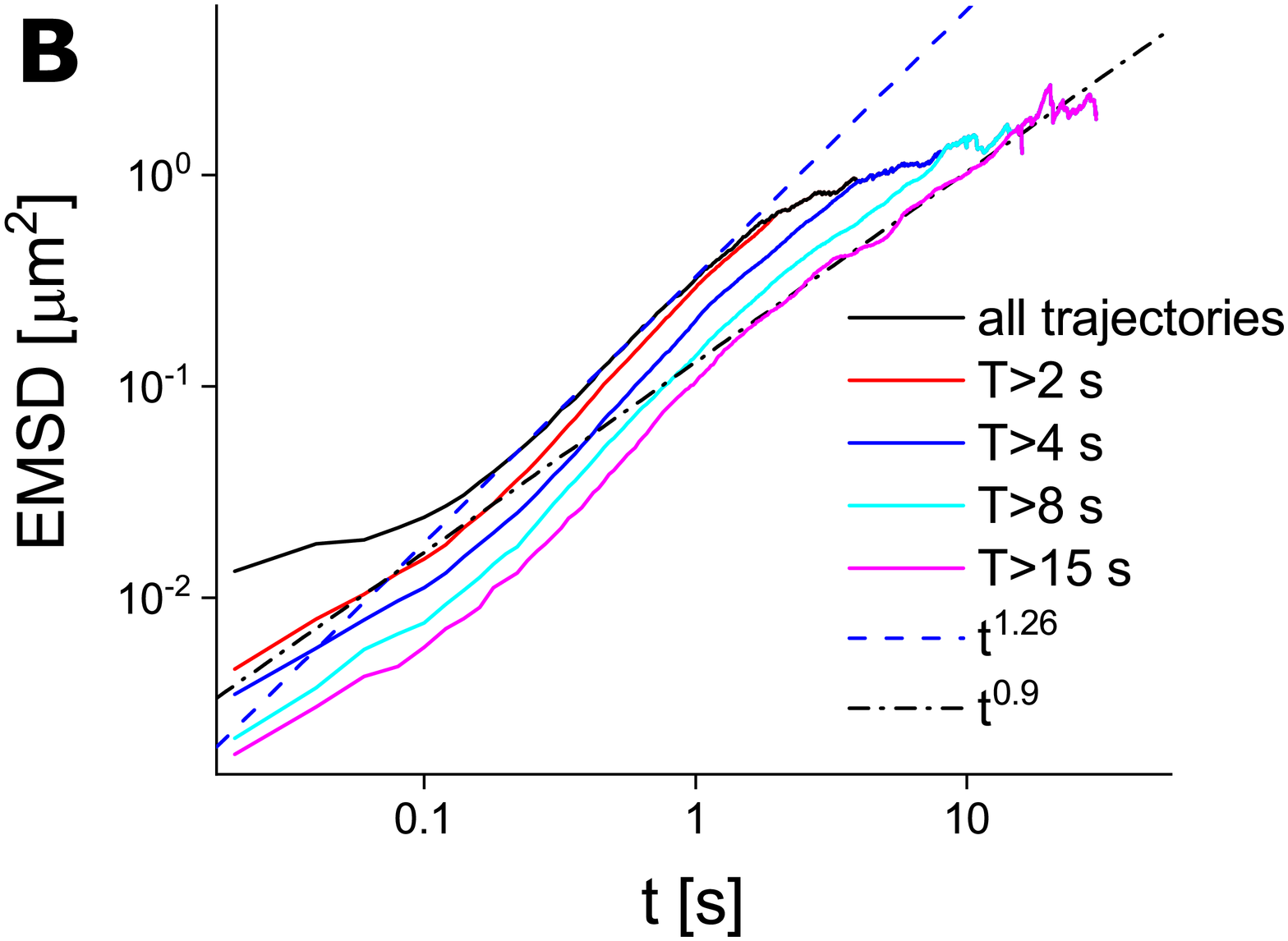}
\includegraphics[scale=0.22]{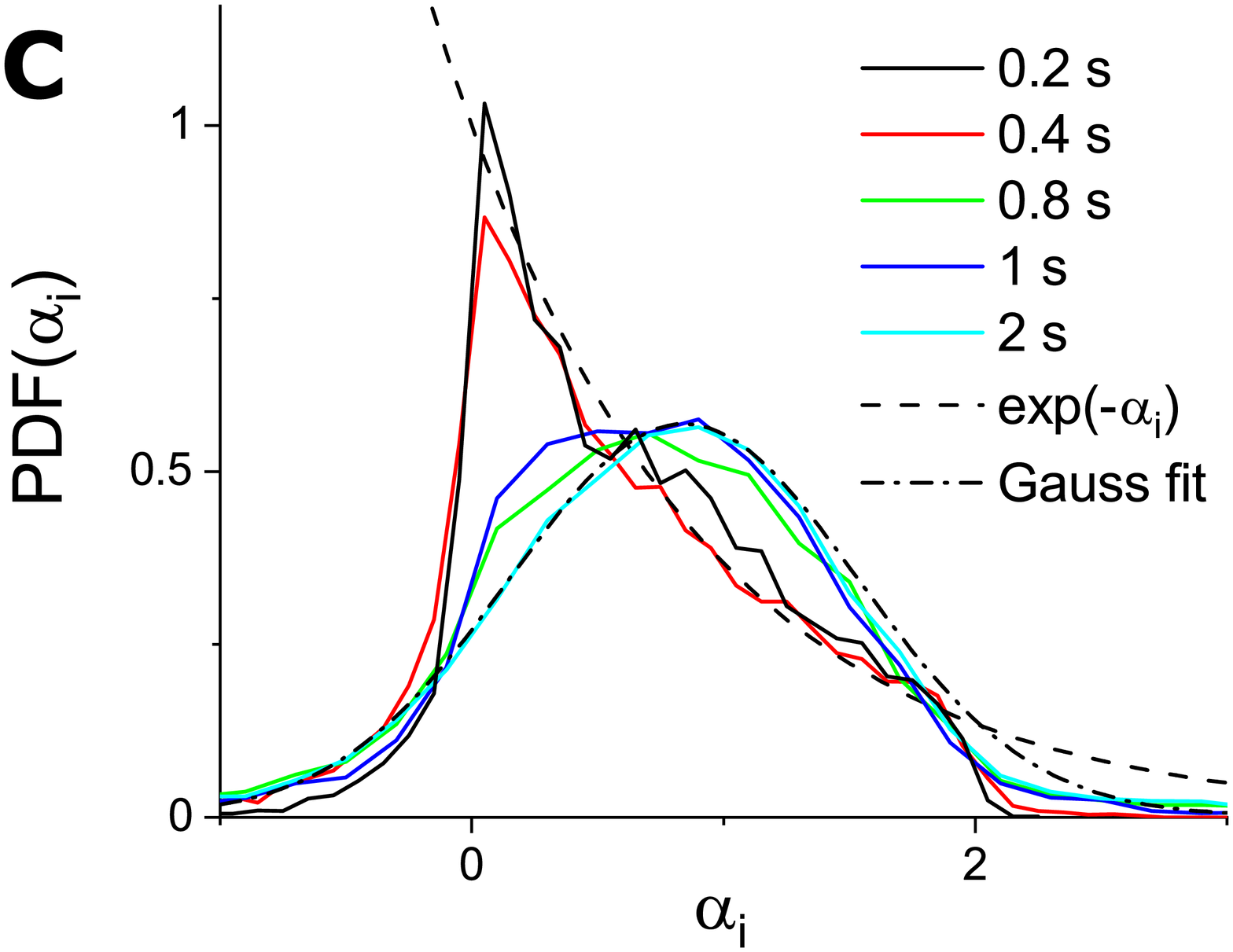}
\includegraphics[scale=0.22]{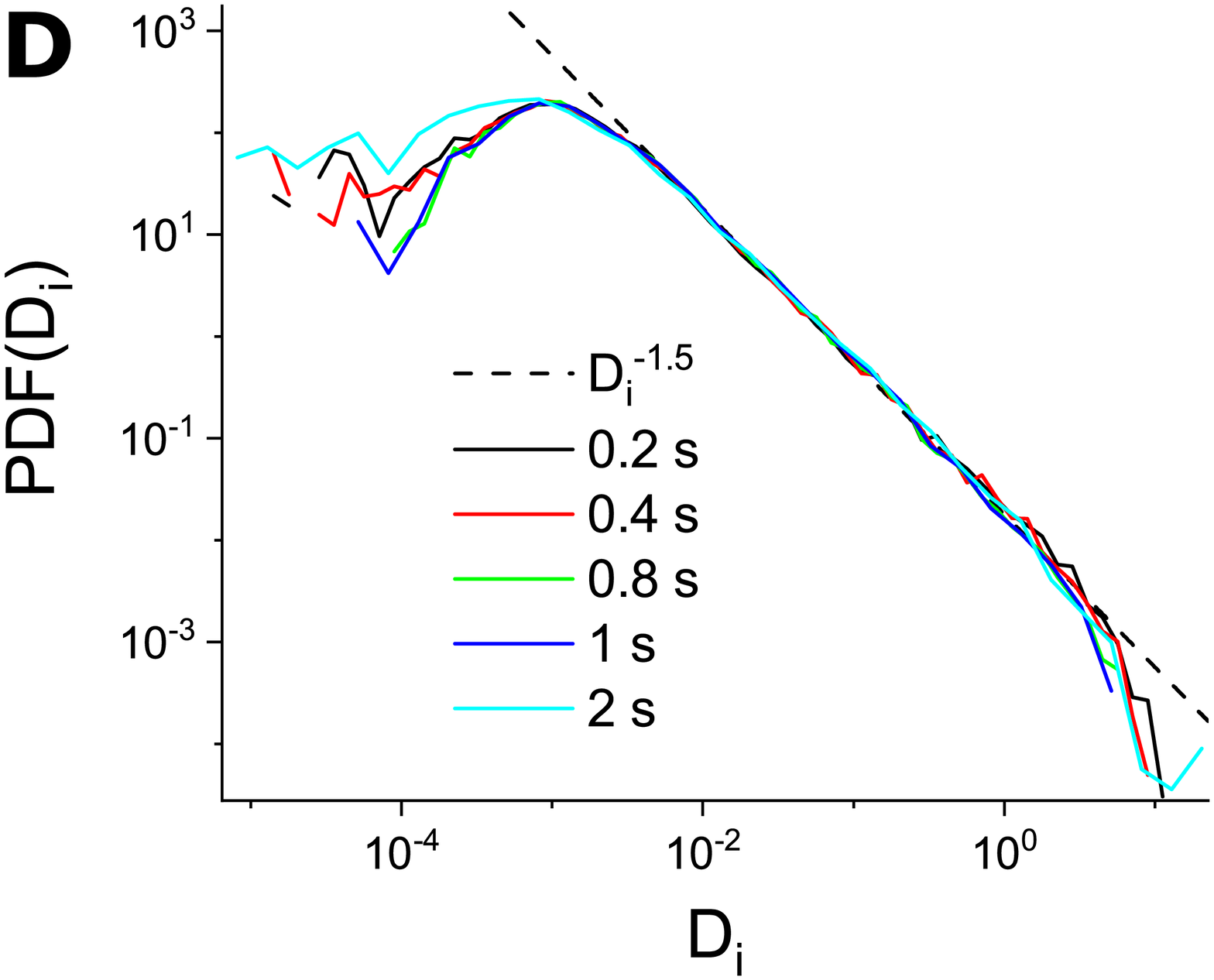}
\caption{\label{EMSD} {\bf EMSD and E-TMSD of experimental trajectories reveal ergodic behaviour. Longer trajectories gradually transition to sub-diffusive behaviour.} {\bf a.} The ensemble-averaged MSD (EMSD) and the time-averaged averaged over all trajectories (E-TMSD) as a function of time interval, t.    
{\bf b.} EMSD as a function of time interval of trajectories longer than $T$ seconds. The super-diffusive behaviour at the intermediate time scale is conserved with the anomalous exponent greater or approximately equal to $\alpha = 1.26 \pm 0.02$. Trajectories longer than $8$ s (curves $T>8$ s and $T>15$ s) become sub-diffusive at longer time scales with the anomalous exponent $\alpha \sim 0.9 \pm 0.01$. {\bf c.} Distributions of anomalous exponents $\alpha_i$ and {\bf d} distributions of generalized diffusion coefficients $D_{\alpha_i}$ obtained by fitting the TMSDs of single trajectories at different times indicated in the legends using the time window $W=0.4$ seconds. Notice that $D_{\alpha_i}$ is dimensionless (see Methods).}
\end{figure*}

From the two-dimensional experimental trajectories  $\mathbf{r}\left(t\right)=\{ x\left(t\right),y\left(t\right)\}$, we extracted the displacements, $X=x\left(t\right)-x\left(0\right)$, $Y=y\left(t\right)-y\left(0\right)$ and displacement increments, $\Delta X=x\left(t+\Delta t\right)-x\left(t\right)$, $\Delta Y=y\left(t+\Delta t\right)-y\left(t\right)$ and calculated their distributions. Endosomal trajectories turned out to be isotropic since the Probability Density Functions (PDFs) of the $X$ and $Y$ components of the displacements are similar. Therefore, we show PDFs of the $X$ component only. The resulting plots are shown in Fig. \ref{PDFX} and \ref{PDFDX}. We find that endosomes display strongly non-Gaussian power-law distributions of displacements and displacements increments along trajectories. The PDFs of scaled displacements $X/\sigma_X$ where $\sigma_X$ is the standard deviation of X, collapse onto a single master curve for $t>0.8$ s (Fig.\ \ref{PDFX}b, c) and are well fitted with the power-law $\left(X/\sigma_X\right)^{-2}$. The form of the PDFs of endosomal displacements does not change in time within the time interval $0.8<t<8$ s. This is different from other observations of heterogeneous systems where it was found that the Laplace distribution of displacement changes its form to a Gaussian distribution at longer times \cite{Granick1,Granick2}. For time $t>8$ s, the PDFs of displacements become noisy due to a small number of remaining trajectories. Distributions of displacement increments along trajectories $\Delta X$ exhibit similar power-law form as shown in Fig.\ \ref{PDFDX}a. Scaled with the standard deviation $\sigma_{\Delta X}$, the PDFs of displacement increments collapse into a single curve, which has a truncated power law behaviour with the same exponent as the PDFs of the displacements (Fig.\ \ref{PDFDX}b, c). Non-Gaussian distributions of displacements and displacement increments are the characteristic feature of heterogeneous diffusion processes (for a recent review see \cite{Metzler2020}). Commonly, Laplace probability distribution of displacements in heterogeneous diffusion is observed, $\mathrm{PDF}\left(X\right)\sim\exp{\left(-\left|X\right|/2\sigma^2\right)}$ \cite{Metzler2020}. For endosomes, we find power-law probability distributions of displacements and increments. To our knowledge, power-law probability distributions were previously reported only for trajectories of individual histone-like nucleoid-structuring proteins \cite{SadoonWang}.

\subsection{Experimental endosomal trajectories display ergodic but anomalous mean-squared displacements}

The ensemble-averaged MSD (EMSD) of experimental trajectories shows the initial sub-diffusive-like behaviour (Fig.\ \ref{EMSD}) which could be attributed to the static measurement errors \cite{Weber}. At time scale $0.2<t<2$ s, molecular motors make the growth of EMSD super-diffusive with the anomalous exponent $\alpha\sim1.26\ \pm0.02$ and the ensemble generalized diffusion coefficient $D_\alpha\sim0.082\pm0.002$ $\mu$m$^2/s^\alpha$. Similar behaviour was observed for the transport of micron-sized beads along microtubules \cite{Caspi} and endosomes filled with labeled nonviral DNA-containing polyplexes ranging from $100$ to $200$ nm in diameter \cite{Kulkarni} and for single cell trajectories \cite{Cherstvy}. At longer time scales ($t\ >\ 2$ s), the EMSD become sub-diffusive (Fig. \ref{EMSD}a) with the anomalous exponent $\alpha \sim 0.5 \pm0.1$. The time averaged MSD along single trajectories (averaged over the ensemble of all trajectories, E-TMSD) is similar to EMSD behaviour. EMSD and E-TMSD are slightly different both in the super-diffusion regime ($t<2$ s) and in the sub-diffusive regime ($t>2$ s), perhaps due to limited statistics. This ergodic behaviour (equivalence of EMSD and E-TMSD) suggests that the endosomal movement is best described by the FBM process. However, the sub-diffusion at long times does not fit the standard FBM model with constant $\alpha$ and $D_\alpha$. As we show below, this apparent sub-diffusion regime is spurious \cite{Etoc}.

\subsection{Time heterogeneous FBM explains spurious sub-diffusive behaviour of endosomal ensemble}

\begin{figure}[ht]
\centering
\includegraphics[scale=0.3]{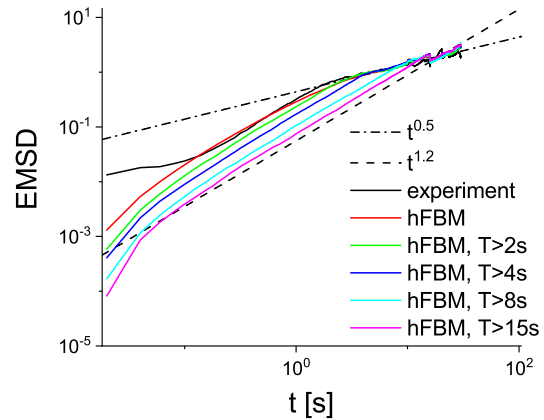}
\caption{{\bf Simulations of ensemble of hFBM trajectories with coupling between diffusion coefficients and duration of trajectories explains the apparent sub-diffusive behaviour of experimental endosome trajectories at longer time scales.} Different curves correspond to EMSDs calculated for simulated hFBM trajectories of duration longer than $T$ seconds. Individual hFBM trajectories were simulated with constant Hurst exponent $H = 0.6$ and constant generalized diffusion coefficients. For standard FBM motion, this corresponds to the super-diffusive growth of EMSD with the anomalous exponent $\alpha = 1.2$ shown by the dashed line. For hFBM, the duration of trajectories was random and the generalized diffusion coefficient for each trajectory was chosen depending on its duration. For longer trajectories, the generalized diffusion coefficient was chosen to be smaller (see Methods for details). The apparent sub-diffusive behaviour at longer time scales with the anomalous exponent $\alpha = 0.5$ is shown as the dashed-dotted line. The EMSD of experimental endosomal trajectories is shown for comparison. 
}
\label{SPUR_SUBDIF}
\end{figure}

To investigate further the behaviour of MSDs, we calculated EMSDs by considering ensembles of experimental trajectories with duration longer than a certain time $T$  (Fig.\ \ref{EMSD}b). The super-diffusive behaviour at the intermediate time scale is conserved with approximately the same anomalous exponent, $\alpha=\ 1.26\ \pm0.02$. For longer trajectories, the generalized diffusion coefficient becomes smaller, which again suggests that slower moving endosomes remain in focus. Notice that the long-time behaviour of EMSD is different from the apparent sub-diffusion observed in Fig.\ \ref{EMSD}a ($\alpha\sim0.5\ \pm0.1$). In fact, for trajectories longer than $T=8$ seconds, EMSDs remain super-diffusive at intermediate time scales much longer and well into the apparent sub-diffusive regime of the EMSD obtained from all trajectories. Moreover, the longtime dynamics gradually transition to a sub-diffusive regime which is different from the sub-diffusive exponent of the EMSD of all trajectories. Trajectories of duration longer than $15$ seconds become sub-diffusive with the exponent $\alpha\sim0.9\ \pm0.05$ (Fig.\ \ref{EMSD}b). The behaviour of the E-TMSDs obtained from trajectories of different duration $T$ is similar to EMSD, therefore they are not shown. To check that the apparent sub-diffusive regime at long time scales could originate from the super-diffusive movement of endosomes, we simulated an ensemble of independent super-diffusive FBM trajectories, each with the same Hurst exponent $H\ =\ 0.6$, but with a different generalized diffusion coefficient. The generalized diffusion coefficients were chosen to be correlated with the time length of trajectory - for longer trajectories we chose smaller generalized diffusion coefficients. This choice is reasonable since in the experimental setup fast endosomes with bigger diffusivities quickly leave the focus area and therefore have shorter duration. On the contrary, slowly moving endosomes with smaller diffusion coefficients stay longer in focus and have longer duration. If it were the case of a standard FBM motion, the EMSD would grow super-diffusively with the anomalous exponents $\alpha=2H=1.2$. Hence, we simulated an ensemble of heterogeneous FBM (hFBM) trajectories of different length $T$. The numerical results are shown in Fig.\ \ref{SPUR_SUBDIF}, confirming that the sub-diffusion regime emerges at long time scales even though all hFBM trajectories were purely super-diffusive. This result shows that time heterogeneous FBM motion explains the spurious sub-diffusion behaviour of MSDs at longer time scales. It also implies that one cannot simply conclude the overall behaviour of the ensemble of endosomes just by looking at the EMSD or the E-TMSD behaviour, since both are masked by the apparent sub-diffusion regime at long time scales caused by the heterogeneity in the trajectory lengths. Therefore, it is necessary to look closer into the local dynamics of individual endosomal trajectories.

\begin{figure*}[ht]
\centering
\includegraphics[scale=0.3]{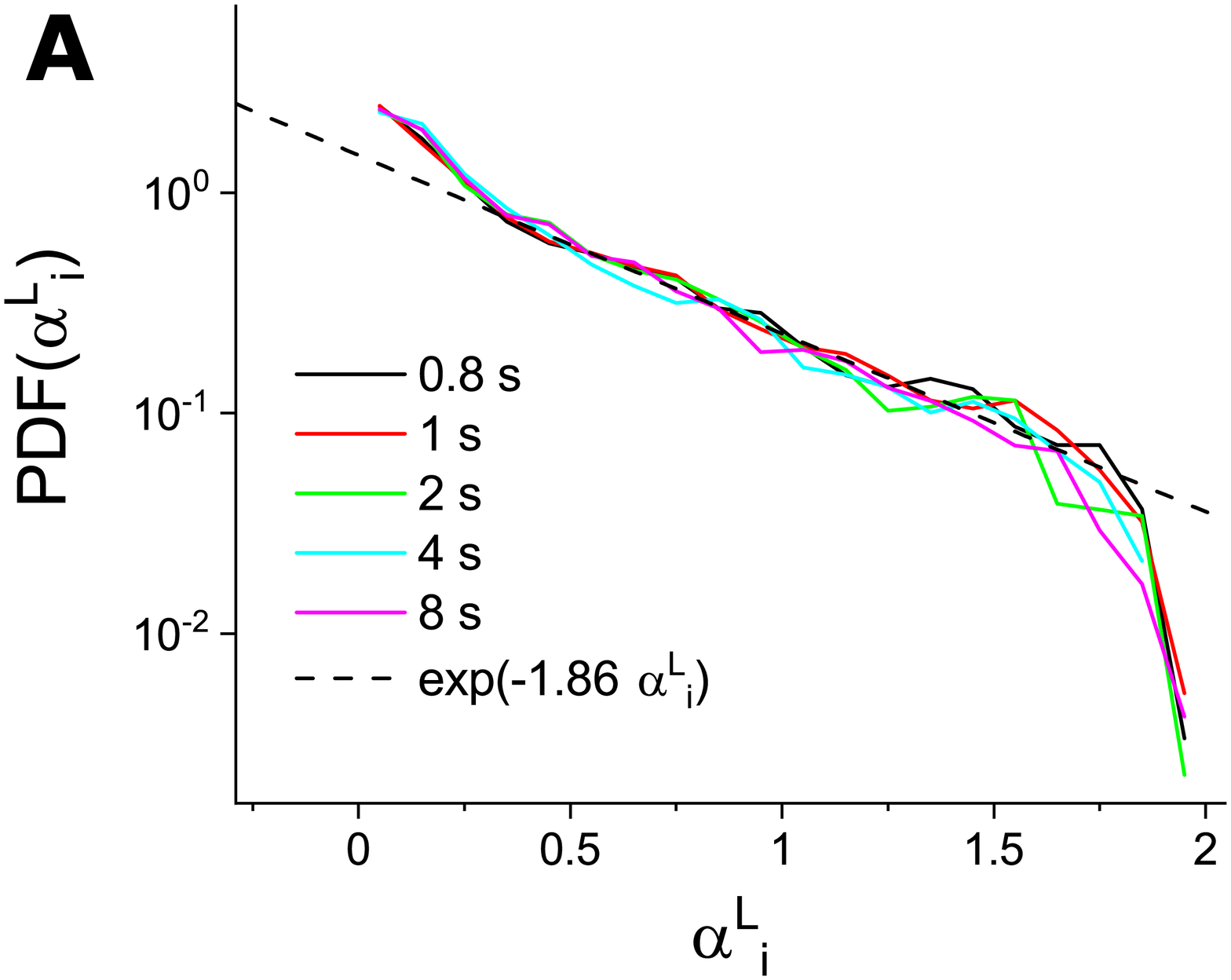}
\includegraphics[scale=0.3]{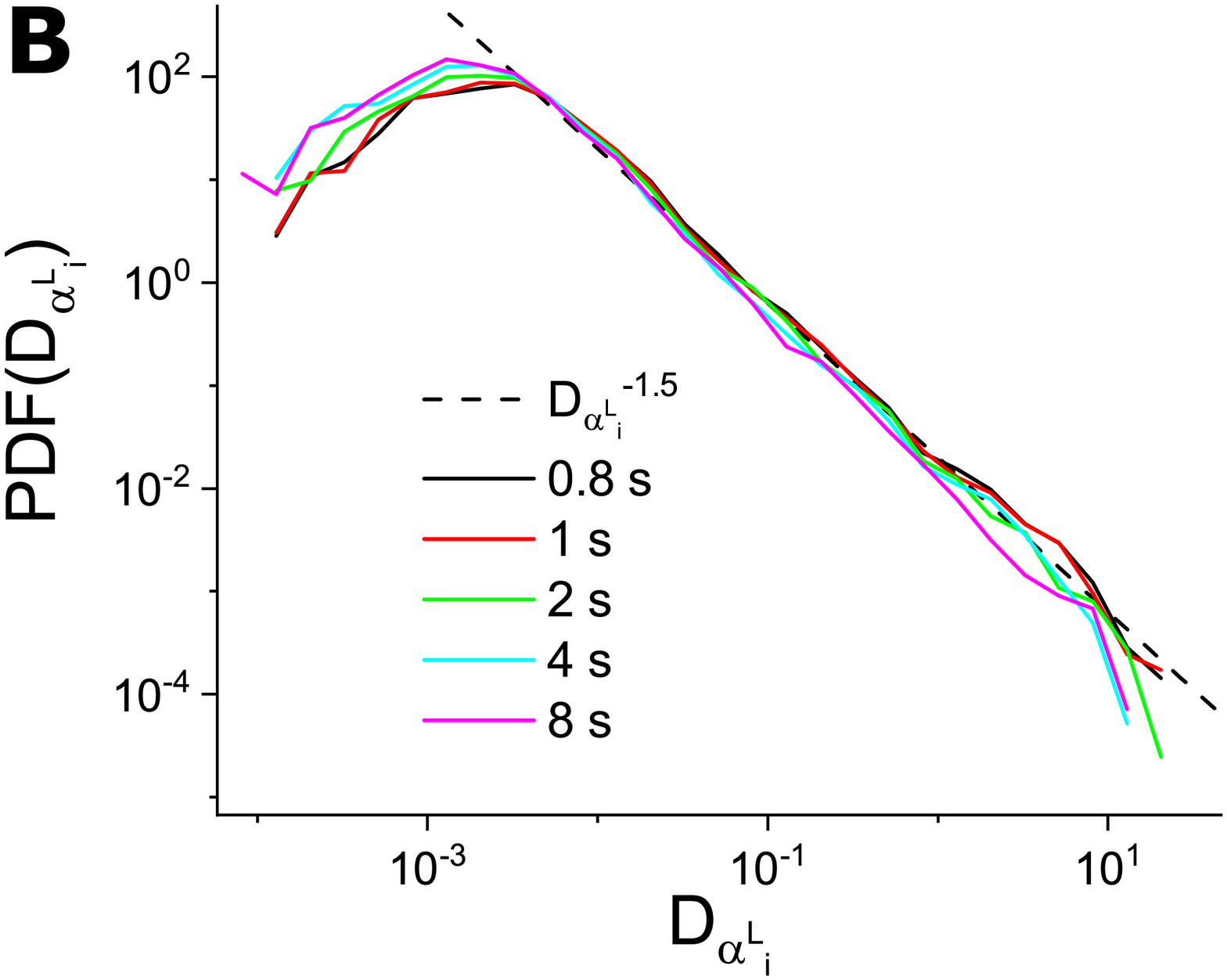}
\caption{{\bf Experimental endosomal trajectories display time independent exponential probability distributions of local anomalous exponents $\alpha^L_i$ and power-law probability distributions of local generalized diffusion coefficients $D_{\alpha^L_i}$.} The local TMSD (L-TMSD) was calculated in a time window $W=1.2$ s at different time $t$. Different curves correspond to PDFs of $\bar{\alpha}$ and $D_{\alpha^L_i}$ calculated at time $t$ given in the figure, see panel {\bf a} and {\bf b} respectively. Notice that $D_{\alpha^L_i}$ is dimensionless (see Methods). The dashed lines represent exponential fit in {\bf a} and power law fit in {\bf b}.}
\label{LOC_ALPHA_D}
\end{figure*}

\begin{figure*}[ht]
\centering
\includegraphics[scale=0.56]{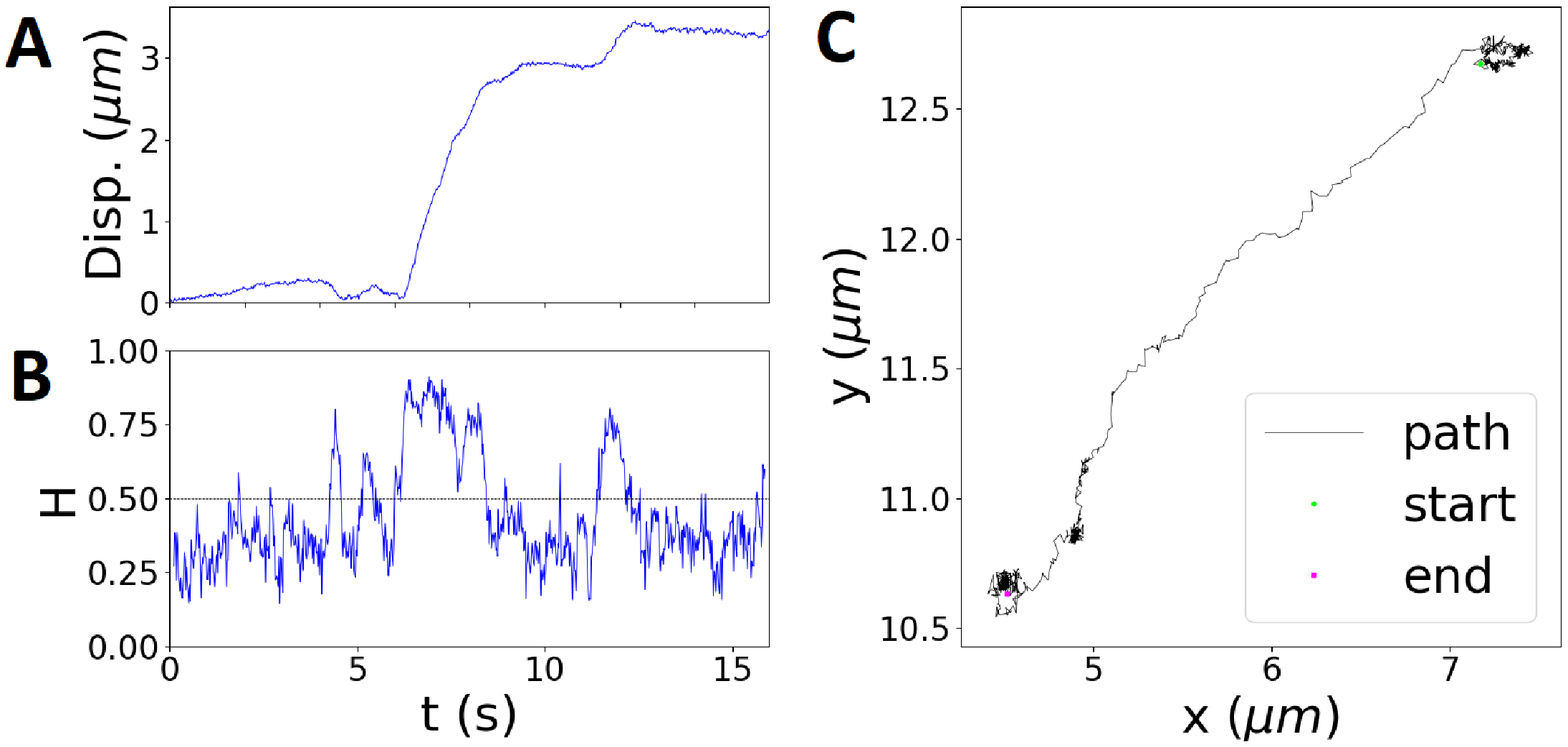}
\includegraphics[scale=0.27]{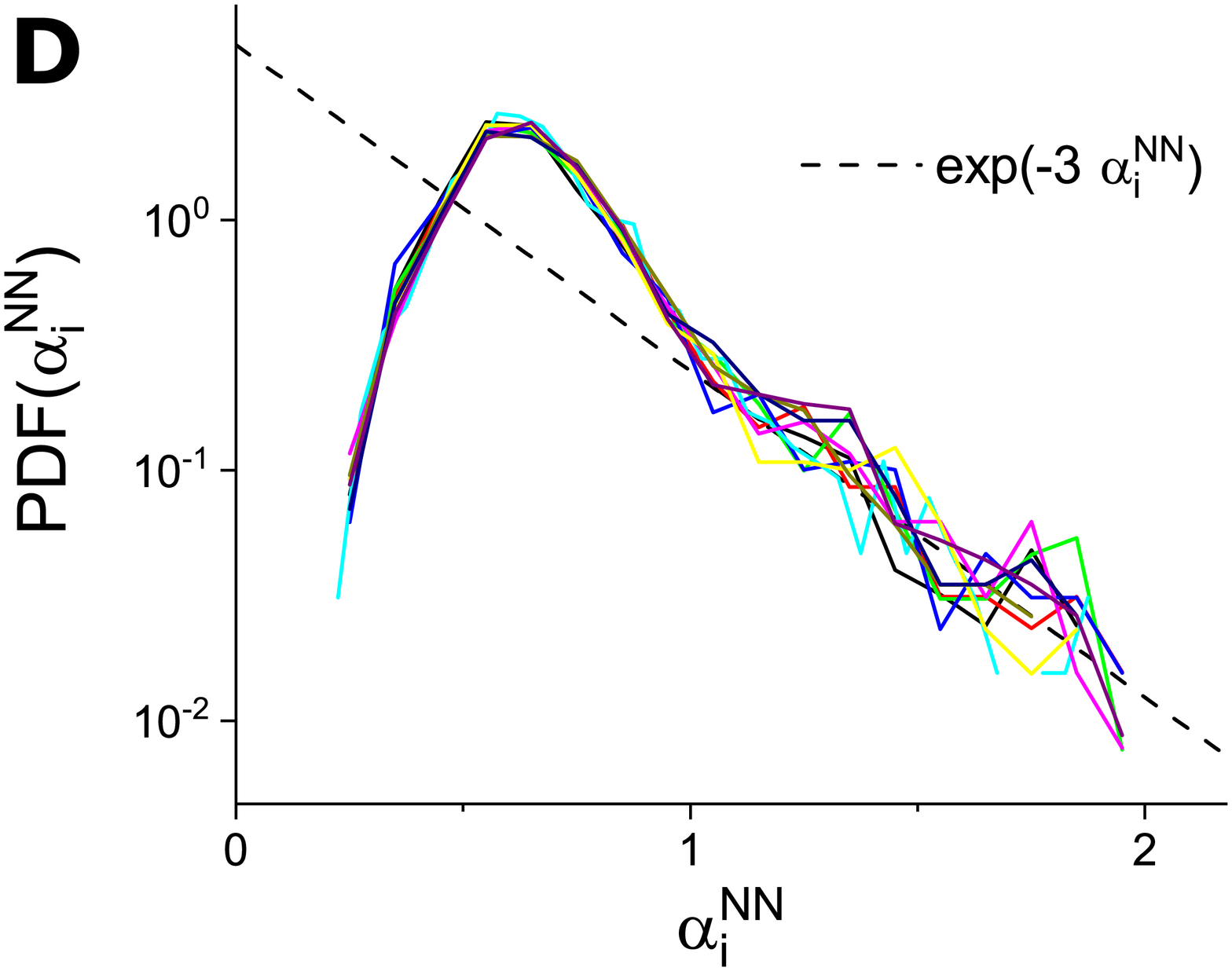}
\hspace{-1.0cm}
\includegraphics[scale=0.27]{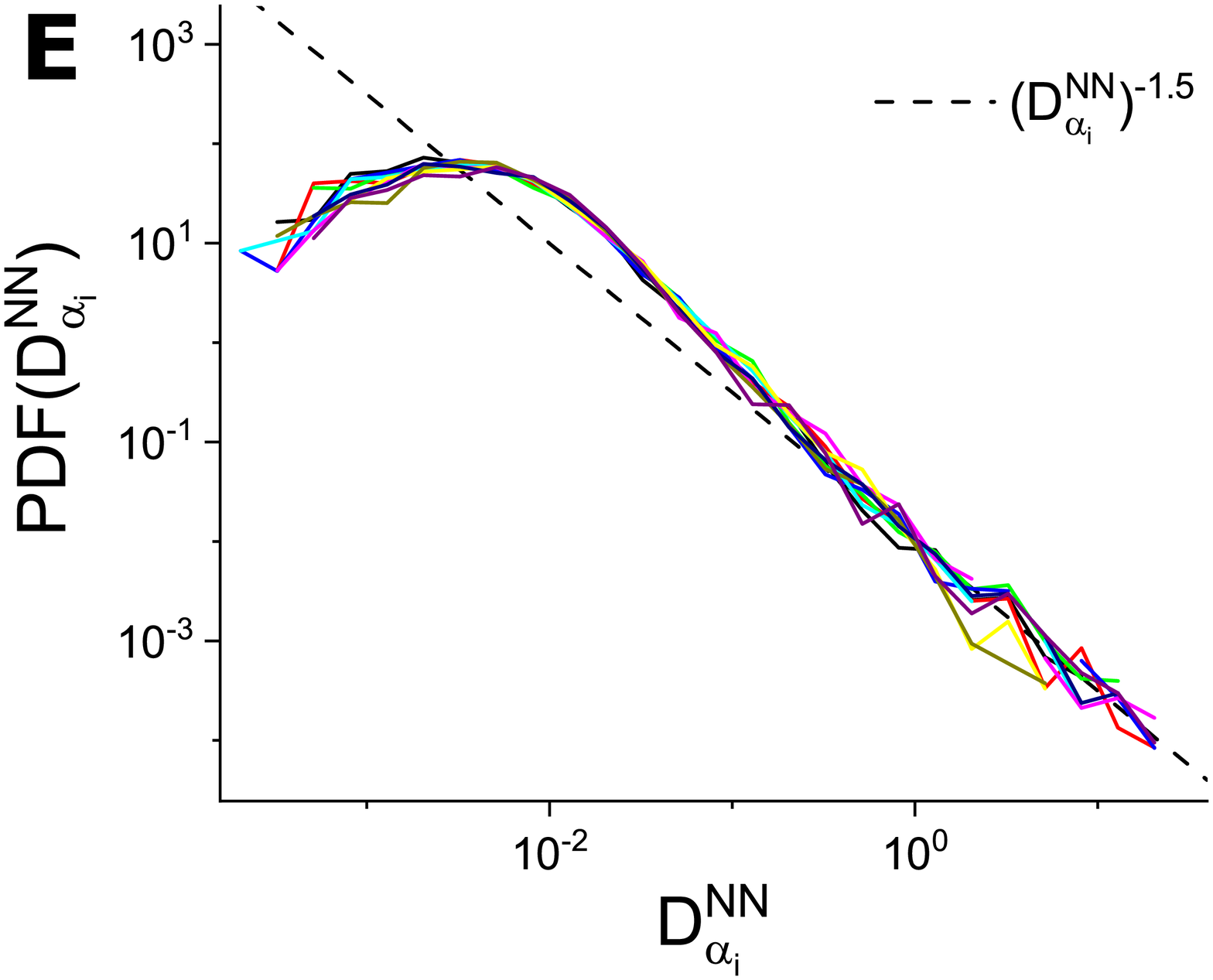}
\caption{{\bf NN analysis reveals spatio-temporal heterogeneity of experimental endosomal trajectories. NN confirms an exponential distribution of anomalous exponent and a power-law distribution of generalized diffusion coefficients.} Displacements $\mathbf{r}(t)-\mathbf{r}(0)$ (an example for a sample trajectory is shown in {\bf a}) were analysed with NN and the local time dependent Hurst exponent $H(t)$ was estimated. {\bf b}. $H(t)$ for a sample trajectory depicted in {\bf c}. {\bf d}. 
Probability density functions of local anomalous exponents $\alpha_{i}^{NN} = 2 H$ estimated with the NN. {\bf e}. PDFs of generalized diffusion coefficients $D_{\alpha^{NN}_i}$. Different curves in {\bf d} and {\bf e} correspond to PDFs estimated at $t = 0.2, 0.4, 0.8, 1, 2, 4, 8, 10, 12, 14$ s.}
\label{H}
\end{figure*}

\subsection{Distributions of time-dependent anomalous exponents and generalised diffusion coefficients}

Using TMSD curves, we calculate the time-dependent anomalous exponent $\alpha_i\left(t\right)$ and the time-dependent generalized diffusion coefficient $D_{\alpha_i}\left(t\right)$ of single experimental endosome trajectories. We did this by calculating TMSDs of single trajectories first, and then fitting TMSDs in a time window $\left(t-W/2,t+W/2\right)$ with the power-law function. From the fit we estimated $\alpha_i\left(t\right)$ and $D_{\alpha_i}\left(t\right)$ (see Methods). The distribution of $\alpha_i\left(t\right)$ (Fig.\ \ref{EMSD}c) is exponential for $t<0.4$ s and (truncated) Gaussian-like at longer time scales. We note that negative exponents $\alpha_i<0$ are fitting artefacts which have no physical meaning. These artefacts arose from the finite duration of experimental trajectories and small window size $W$. For larger window size, the fitting artefacts gradually diminish (see Appendix B, Fig. S2). The peak of the Gaussian fit (dashed-dotted curve in (Fig.\ \ref{EMSD}c) corresponds to $\alpha_i\sim0.9$ which agrees with the EMSD behaviour at this time scale (Fig.\ \ref{EMSD}b). PDFs of $D_{\alpha_i}$ (Fig.\ \ref{EMSD}d) are independent of time and are best fitted with a power law (shown in Fig.\ \ref{EMSD}d as the dashed line). In contrast to the constant ensemble anomalous exponent $\alpha$ and the constant generalized diffusion coefficient $D_\alpha$, anomalous exponents $\alpha_i\left(t\right)$ and generalized diffusion coefficients $D_{\alpha_i}\left(t\right)$ characterize endosomal motion in time, but also in space. Indeed, $\alpha_i\left(t\right)$ and $D_{\alpha_i}\left(t\right)$ contain some amount of spatial averaging since they were calculated within a finite time window of the TMSD curves, obtained collecting displacements at the time scale $t$ from any point in the endosomal trajectory (see the Methods for the TMSD definition). On top of this, we already noticed that the distributions of the time-dependent anomalous exponents $\alpha_i\left(t\right)$ have fitting artefacts. Therefore, in the next subsection we proceed to examine the local-time characteristics of individual trajectories by estimating the anomalous exponent and generalized diffusion coefficient straight from endosomal trajectories.

\subsection{Endosomes display exponential distributions of local anomalous exponents and power law distributions of generalised diffusion coefficients}

To examine the local characteristics of individual trajectories, we calculated the local TMSDs (L-TMSD) from each experimental trajectory at the time $t$. This is defined as the TMSD of a single trajectory's portion contained within the time window $\left(t-W/2,t+W/2\right)$ \cite{Heinrich}. L-TMSDs were fitted with a power law function (see Methods) and the local anomalous exponents $\alpha_i^L\left(t\right)$ and the local generalized diffusion coefficients $D_{\alpha_i^L}\left(t\right)$ were estimated. In general, $\alpha_i^L$ and $D_{\alpha_i^L}$ are different from $\alpha_i$ and $D_{\alpha_i}$. On the time scale $0.2\ <\ t\ <\ 8$ s, the PDFs of $\alpha_i^L$ and $D_{\alpha_i^L}$ do not depend on time (Fig.\ \ref{LOC_ALPHA_D}a, b). They also do not depend on window size (see Appendix B, Fig. S3). Most importantly, $\alpha_i^L$ and $D_{\alpha_i^L}$ of experimental endosomal trajectories are best fitted with exponential and power law functions respectively (Fig.\ \ref{LOC_ALPHA_D}a, b). Averaged over all trajectories, the values of $\alpha_i^L$ and $D_{\alpha_i^L}$ at different times are markedly different from the values of $\alpha$ and $D_\alpha$ estimated from the EMSDs (see Appendix B, Fig. S4). PDFs of local anomalous exponents $\alpha_i^L$ (Fig.\ \ref{LOC_ALPHA_D}a) show that despite super-diffusive motion being present at all times, the average local anomalous exponent $\alpha_i^L$ is sub-diffusive. The values of the local anomalous exponents and generalized diffusion coefficients exhibit non-linear positive correlations (see Appendix B, Fig. S5) ($\alpha_i^L=a\log{D_{\alpha_i^L}}+b$ with $a=0.42$, $b=1.22$ and Pearson's correlation coefficient $r=0.8$).  

\begin{figure*}[ht]
\centering
\includegraphics[scale=0.3]{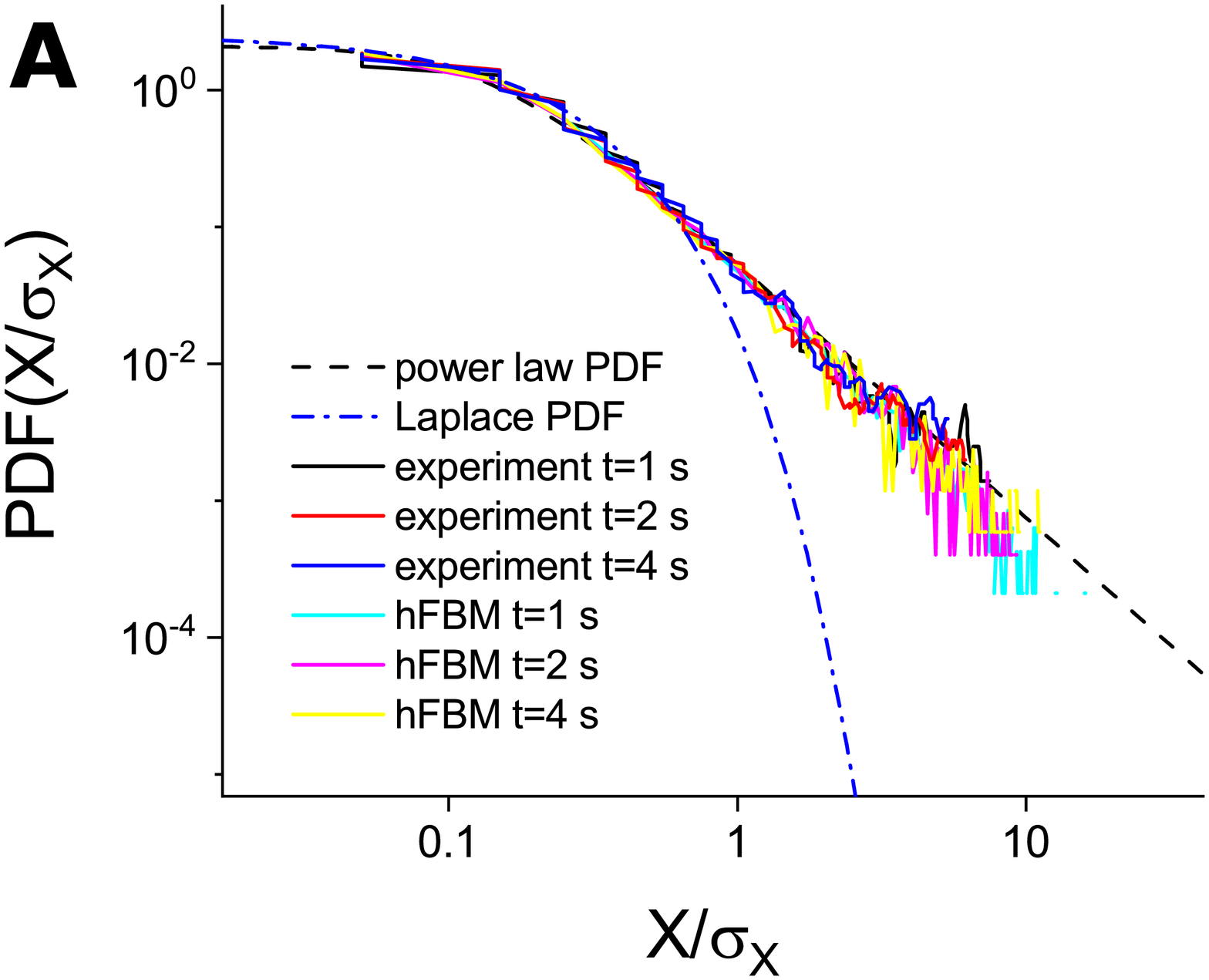}
\includegraphics[scale=0.3]{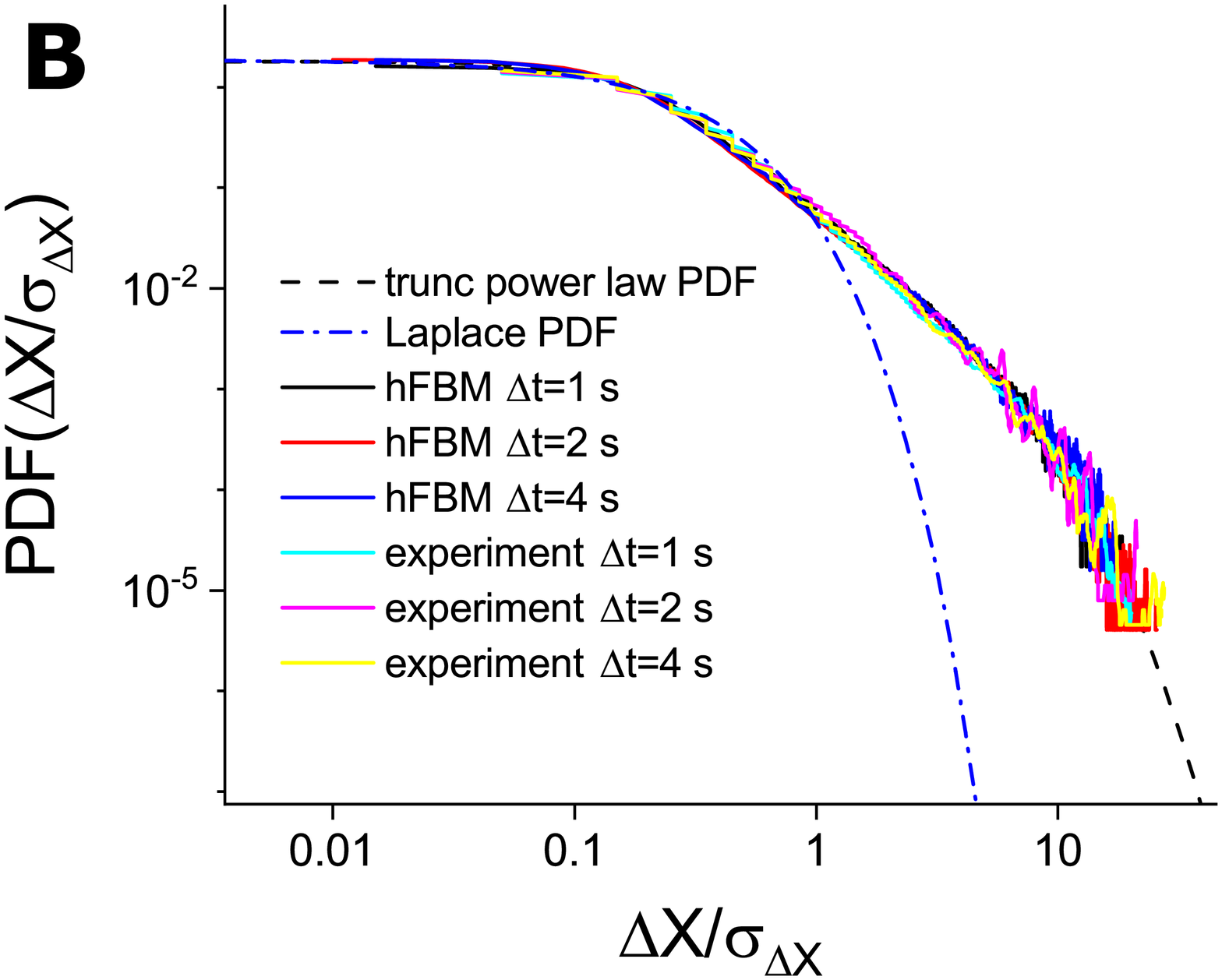}
\caption{{\bf Power-law probability distributions of displacements and increments of experimental endosomal trajectories are the same as for the heterogeneous ensemble of FBM trajectories.} Probability distributions of displacements $X=x(t)-x(0)$ scaled by the standard deviation $\sigma_X$ {\bf a} and increments $\Delta X=x(t+\Delta t)-x(t)$ scaled by the standard deviation $\sigma_{\Delta X}$ {\bf b} for the hFBM trajectories with the generalized diffusion coefficients distributed according to a power law. The PDFs of displacements and increments for experimental trajectories overlap with PDFs for hFBM. The dashed and dashed-dotted curves are the fits with the power law {\bf a}, truncated power law {\bf b} and Laplace PDFs.}
\label{hFBM}
\end{figure*}

To assess the local characteristics of a single endosome motion using a different method, we applied the neural network (NN) \cite{ELife}. Using NN we estimated the local Hurst exponent $H$ of experimental trajectories which is related to the local anomalous exponent, $H=\alpha/2$. The advantage of the NN is that it does not make use the fit of TMSD to estimate $H$ and it is a more sensitive estimator \cite{ELife}. Since endosomes MSD appears to be ergodic (see Fig.\ \ref{EMSD}a), we trained the NN on FBM trajectories and estimated the local Hurst exponents $H\left(t\right)$ for each endosomal trajectory (see \cite{ELife} for details of the implementation and training of the NN). An example of a heterogeneous endosomal trajectory and the local Hurst exponent estimated using the NN is shown in Fig.\ \ref{H}a-c. The figure shows that the endosome movement switches between slow anti-persistent ($H<0.5$) and fast persistent movements $H>0.5$. The PDFs of $\alpha_i^{NN}$ estimated using the NN (Fig.\ \ref{H}d) are qualitatively similar to the distributions of the local anomalous exponents (Fig.\ \ref{LOC_ALPHA_D}a) although the exponential decay of PDFs of $\alpha_i^{NN}$ is faster. Moreover, the PDFs of $\alpha_i^{NN}$ display maxima around $\alpha_i^{NN}\sim0.7$. We stress the fact that, although the power law fitting of L-TMSD may produce non-physical values of $\alpha_i^L$ ($\alpha_i^L<0$ or $\alpha_i^L>2$), the estimate of $\alpha_i^{NN}$ is free of this drawback. The NN methods could directly infer only the value of the anomalous exponent. For the estimate of the generalized diffusion coefficient $D_{\alpha_i^{NN}}$ we proceeded as follows. Handling the same endosome trajectories portions used in the L-TMSD analysis, we fit the resulting MSD with the law $D_{\alpha_i^{NN}} \; t^{\alpha_i^{NN}}$. As a consequence, $\alpha_i^{NN}$ and $D_{\alpha_i^{NN}}$ are non-linearly positively correlated as $\log{\alpha_i^{NN}}=c\log{D_{\alpha_i^{NN}}}+d$ with $c=0.03$, $d=0.12$ and the Pearson's correlation coefficient $r=0.54$ (see Supplementary Note, Fig. S4). Importantly, by comparing the distributions of $D_{\alpha_i^L}$ and $D_{\alpha_i^{NN}}$ (Figs.\ \ref{LOC_ALPHA_D}b, \ref{H}e), we can conclude that the power-law decay appears robust to the specific methodology used. At the same time, stressing the difference among the PDFs of the local anomalous exponents in Fig. \ref{LOC_ALPHA_D}a and Fig. \ref{H}d and the inherent fitting problems, we can assert that the NN method is more reliable.

\begin{figure*}[ht]
\centering
\includegraphics[scale=0.3]{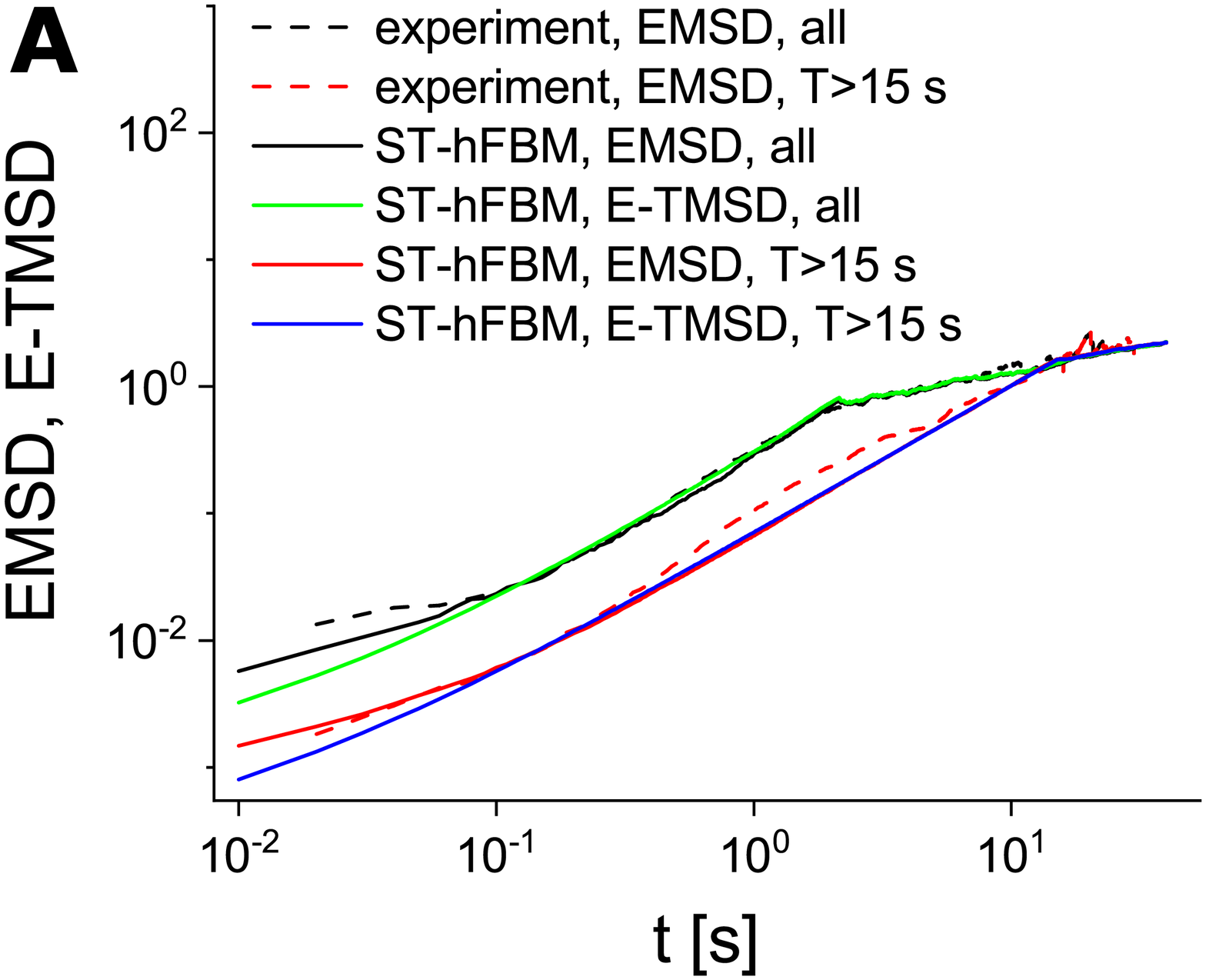}
\includegraphics[scale=0.3]{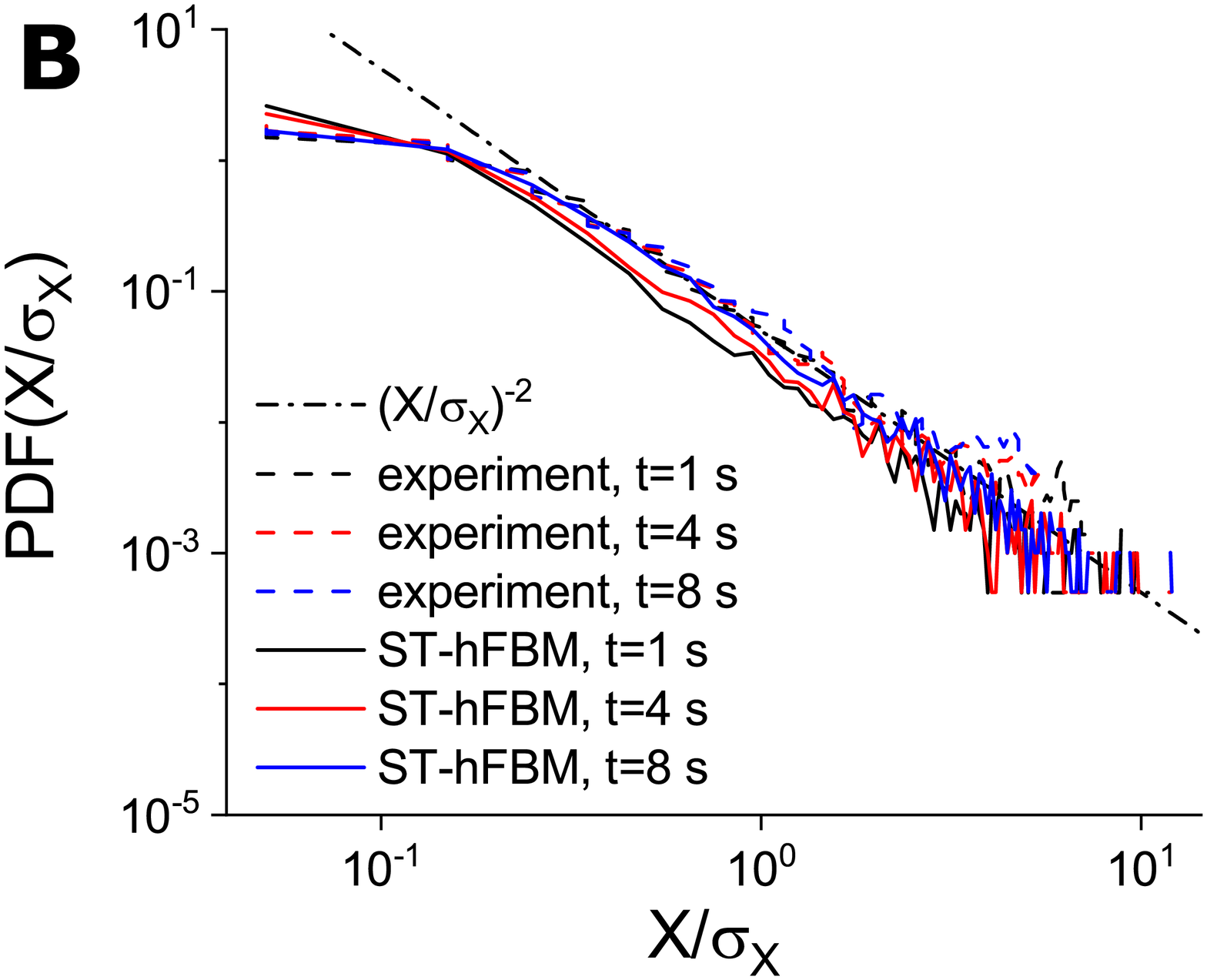}
\includegraphics[scale=0.3]{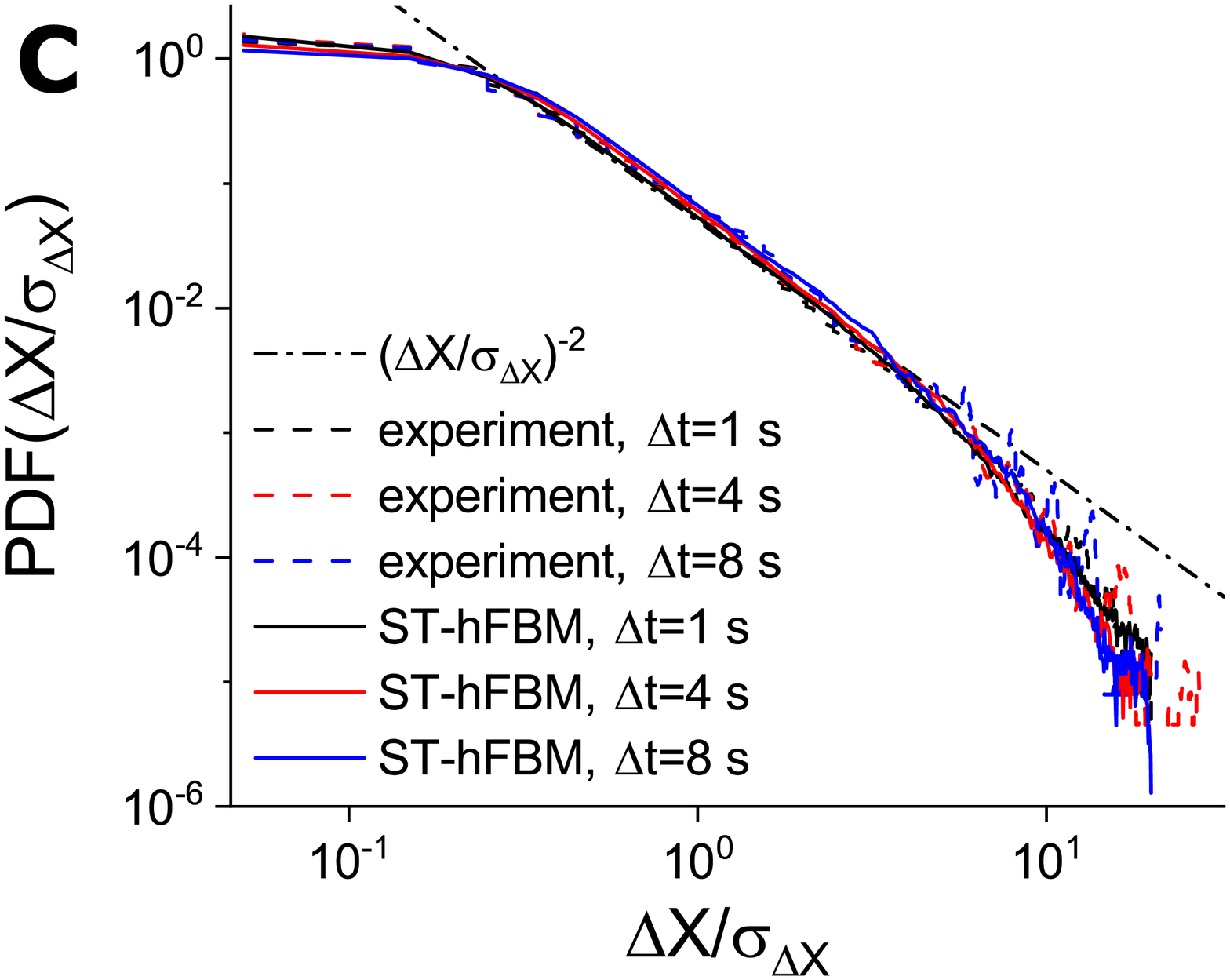}
\caption{{\bf Spatio-temportal heterogeneous FBM model (ST-hFBM) explains MSDs, power-law probability distributions of displacements and increments of experimental endosomal trajectories.} {\bf a} EMSD and E-TMSD of ST-hFBM trajectories compared with the EMSD of experimental trajectories. Different curves correspond to different duration of trajectories $T$ used to calculate EMSD. {\bf b} Probability distributions of displacements $X=x(t)-x(0)$ scaled by the standard deviation $\sigma_X$ and {\bf c} increments $\Delta X=x(t+\Delta t)-x(t)$ scaled by the standard deviation $\sigma_{\Delta X}$ for the ST-hFBM trajectories compared with the corresponding PDFs of experimental trajectories.}
\label{SThFBM}
\end{figure*}

\subsection{Endosome movement is a spatio-temporal heterogeneous fractional Brownian motion}

The ergodic behaviour of the anomalous MSDs of experimental endosomal trajectories (Fig.\ \ref{EMSD}a) and the stationary (time independent) property of their scaled by the standard deviation displacements and increments distributions (Figs.\ \ref{PDFX}b,c and \ref{PDFDX}b,c) agree with the FBM model. On the other hand, the power law form of distributions of displacements and increments and the PDFs of local anomalous exponents and generalized diffusion coefficients are not captured by the standard FBM. Below we show that heterogeneous FBM (hfBm) explains all the experimental data for endosomal transport in eukaryotic cells.

\subsubsection{Spatially heterogeneous fractional Brownian motion explains PDFs of displacements and increments, but fails to capture MSDs of the endosome ensemble}

A simple form of heterogeneous FBM behaviour which is only spatially (but not temporally) heterogeneous could be realised via an ensemble of independent FBM trajectories with Hurst exponent $H$ and generalized diffusion coefficient $D_H$, different for each trajectory and constant in time \cite{itto-beck-2021}. $H$ was drawn from an exponential probability distribution, while $D_H$ was distributed in a power law fashion in analogy to Fig.\ \ref{LOC_ALPHA_D} and Fig. \ref{H}d,e (see Methods for details). This simplified picture corresponds to the spatial inhomogeneity of FBM motion, while no time heterogeneity was implemented. Figure \ref{hFBM} demonstrates an excellent agreement between distributions of displacements and increments of the spatially heterogeneous ensemble of FBM trajectories and PDFs obtained from experimental trajectories. Analytical calculations (see Appendix A and Appendix B, Fig. S6) confirm the existence of the scaling behaviour (collapse of PDFs on a master curve) of the PDFs of displacements of hFBM. The velocity autocorrelation functions of the ensemble of hFBM trajectories also show similar qualitative behaviour compared to the experimental trajectories (see Appendix B, Fig. S6). However, the EMSD of the hFBM ensemble does not agree with the EMSD of experimental trajectories (see Appendix A, Eq. (8) for analytical derivation). This discrepancy arises because of the simplified picture of the spatially heterogeneous FBM approach which completely disregards the trajectories' time-heterogeneity. Therefore, we ought to consider a spatio-temporal heterogeneous FBM.

\subsubsection{A spatio-temporal heterogenous FBM best describes experimental endosome motion}

Opposite to the spatially heterogeneous FBM, the Hurst exponent of a single spatio-temporal heterogeneous FBM (ST-hFBM) trajectory is not constant but randomly switches in time between anti-persistent $H=0.25$ value and persistent $H=0.6$ value (see Methods). Such a dichotomous model corresponds to random switching of endosomal movement between active and passive states of motion as observed in experiment (see Fig.\ \ref{H}b). Results of simulations of ST-hFBM qualitatively agree with the experimental data (Fig.\ \ref{SThFBM}). The EMSDs, distributions of displacements and increments of simulated ST-hFBM trajectories fit well the EMSDs and corresponding distributions of experimental trajectories. This supports our assumption that the endosomal motion is best described by a spatio-temporal heterogeneous FBM. The quantitative deviations suggest that there could be more than two states in endosome motion. The multiple active states most probably will correspond to endosomal movement powered with different number/type of molecular motors \cite{PLOS,Fedotov,Kenwright}. 

In summary, our results show that including spatio-temporal heterogeneity into the FBM captures essential statistical characteristics of endosomal motion (Figs. \ref{PDFX}, \ref{PDFDX}, \ref{EMSD}). The analysis of local dynamic properties reveals a robust character of endosomal dynamics and a broad spectrum of local anomalous exponents and generalized diffusion coefficients. The distributions of local anomalous exponents and generalized diffusion coefficients do not depend on time and best described by the exponential and power-law PDFs respectively (Figs. \ref{LOC_ALPHA_D}, \ref{H}). The dependence of the generalized diffusion coefficient on trajectory duration in the hFBM also explains the apparent sub-diffusive behaviour seen at long-time scales (Fig. \ref{SPUR_SUBDIF}). Moreover, a simple spatially heterogeneous FBM accurately reproduces the power-law distributions of displacements and increments of endosomes (Fig. \ref{hFBM}). By incorporating both the spatial and temporal heterogeneity found in experiments (Fig. \ref{H}), we show that the ST-hFBM describes experimental endosomal trajectories. It reproduces the MSDs (including the ergodic behaviour in terms of the equivalence of EMSD and E-TMSD), displacement and increment probability distributions of endosomes (Fig. \ref{SThFBM}).

\section{Methods}

\subsection{Statistical analysis of experimental trajectories} 

\subsubsection{EMSD, $\alpha$ and $D_{\alpha}$} 

The ensemble-averaged mean squared displacement (EMSD) of 2D experimental trajectories is defined in a standard way:
\begin{equation}
\mbox{EMSD}(t) = \frac{\left< \mathbf{r} \right>^2(t)}{l^2},
\label{emsd}
\end{equation}
where $l$ is the length scale which we choose $l=1$ $\mu$m and
\begin{equation}
\left< \mathbf{r} \right>^2(t) = \left< (x_i(t)-x_i(0))^2 + (y_i(t)-y_i(0))^2 \right>,
\label{emsd1}
\end{equation}
The angled brackets denotes averaging over an ensemble of trajectories, $\left< A \right>=\sum_{i=1}^{N} A_i/N$, where $N$ is the number of trajectories in the ensemble. 

By fitting the EMSD to the power law function, we extracted the anomalous exponent $\alpha$ and the generalized diffusion coefficient $D_{\alpha}$ 
\begin{equation}
\mbox{EMSD}(t) = 4 D_{\alpha} \left( \frac{t}{\tau} \right)^{\alpha}.
\label{emsdfit}
\end{equation}
The time scale $\tau=1$ sec and the length scale $l=1$ $\mu$m are introduced in order to make the generalized diffusion coefficient $D_{\alpha}$ dimensionless \cite{Heinrich2015}. Notice that $\alpha$ and $D_{\alpha}$ are constants which characterize averaged transport properties of ensemble of endosomal trajectories. 

\subsubsection{TMSD, E-TMSD, $\alpha_i(t)$ and $D_{\alpha_i}(t)$} 

The subscript $i$ indicates that $\alpha_i(t)$ and $D_{\alpha_i}(t)$ are extracted for an individual trajectory with index $i$ as described below. $\alpha_i(t)$ and $D_{\alpha_i}(t)$ are also calculated at time $t$ (so, they depend on time) in contrast to $\alpha$ and $D_{\alpha}$ which are constants calculated for ensemble of all trajectories (see the definition above). First, we calculate the time-averaged mean squared displacement (TMSD) of a individual trajectory $\{x_i, y_i \}$ of a duration $T$:
\begin{equation}
\mbox{TMSD}_i(t) = \frac{\overline{\delta^2(t)}}{l^2}, 
\label{tmsd}
\end{equation}
where where $l$ is the length scale which we choose $l=1$ $\mu$m and
\begin{equation}
\overline{\delta^2(t)} = \frac{ \int_{0}^{T-t} \Delta_i(t,t') dt'}{T-t},
\label{tmsd1}
\end{equation}
where
\begin{equation}
\Delta_i(t,t') = (x_i(t'+t)-x_i(t'))^2 + (y_i(t'+t)-y_i(t'))^2.
\end{equation}
TMSDs of individual trajectories can be averaged further over the ensemble of trajectories to get the ensemble-time-averaged MSD (E-TMSD): 
\begin{equation}
\mbox{E-TMSD}(t) = \left< \mbox{TMSD}_i(t) \right>.
\label{etmsd}
\end{equation}
By fitting the TMSDs to the power law function in a time window $(t-W/2,t+W/2)$, we extract the anomalous exponent $\alpha_i(t)$ and the generalized diffusion coefficient $D_{\alpha_i}(t)$ at time $t$:
\begin{equation}
\mbox{TMSD}_i(t) = 4 D_{\alpha_i} \left( \frac{t}{\tau} \right)^{\alpha_i}, \; \; \;  (t-W/2,t+W/2).
\label{tmsdfit}
\end{equation}

\subsubsection{Local TMSD (L-TMSD), $\alpha^L_i(t)$ and $D_{\alpha^L_i}(t)$} 

We calculate the local TMSD (L-TMSD) e.g. TMSD is calculated in a time window $(t-W/2,t+W/2)$. The window is then shifted along a trajectory and L-TMSD is calculated in each window. The subscript $i$ indicates that $\alpha^L_i(t)$ and $D_{\alpha^L_i}(t)$ are extracted for an individual trajectory with index $i$. By fitting the first $10$ L-TMSDs points to power law functions, we extracted the time dependent local anomalous exponents $\bar{\alpha}(t)$ and the time dependent local generalized diffusion coefficients $D_{\bar{\alpha}}(t)$:
\begin{equation}
\mbox{L-TMSD}_i(t) = 4 D_{\alpha^L_i} \left( \frac{t}{\tau} \right)^{\alpha^L_i}.
\label{ltmsdfit}
\end{equation}
Here we introduced the time scale $\tau=1$ sec and the length scale $l=1$ $\mu$m in order to make the local generalized diffusion coefficients dimensionless \cite{Heinrich2015}. 
Notice the difference between the TMSD fit in Eq.\ \ref{tmsdfit} and the L-TMSD fit in Eq.\ \ref{ltmsdfit}. In Eq.\ \ref{tmsdfit} we calculate the TMSD first and then fit it in the time window $(t-W/2,t+W/2)$. In Eq.\ \ref{ltmsdfit} we first calculate the L-TMSD (TMSD in the time window $(t-W/2,t+W/2)$) and then fit the first $10$ points of L-TMSD. 

The time dependence of the EMSD determine the character of the movement of ensemble of endosomes as normal diffusion for $\alpha=1$, sub-diffusion for $\alpha<1$ and super-diffusion for $\alpha>1$. Similarly the time dependence of the L-TMSD determine the character of local movement of a single endosome as normal diffusion for $\alpha^L_i=1$, sub-diffusion for $\alpha^L_i<1$ and super-diffusion for $\alpha^L_i>1$.

\subsection{ Simulation of spatially heterogeneous FBM trajectories (Fig.\ \ref{SPUR_SUBDIF})}

FBM trajectories were simulated in MATLAB (The MathWorks, Nantick, MA) ffgn function which uses the Fast Fourier Transform method to generate FBM. Each trajectory is given a constant Hurst exponent, $H$, and generalized diffusion coefficient, $D$ \cite{itto-beck-2021}. In simulations shown in Fig.\ \ref{SPUR_SUBDIF}, the Hurst exponents were constant ($H=0.6$) for all trajectories. The generalized diffusion coefficients were assigned from the power law distribution $f(D)=\gamma D_{min}^{\gamma} D^{-1-\gamma}$ with $\gamma=0.5$ as it was observed for endosomal trajectories (see Fig.\ \ref{LOC_ALPHA_D}b). The PDF of $D$ was truncated at $D_{min}=0.001$ and $D_{max}=7$. To reproduce the sub-diffusive behaviour of MSDs at longer time scales, the duration of the hFBM trajectories $T$ was coupled with their generalized diffusion coefficients. Specifically, we chose $T = 5 D^{-0.6}$ s which was truncated by $T_{min}=0.04$ s and $T_{max}=70$ s. So, longer trajectories have smaller generalized diffusion coefficient $D$.

\subsection{ Simulation of spatially heterogeneous FBM trajectories (Fig.\ \ref{hFBM})}

FBM trajectories were simulated with the same method as detailed above. In contrast with simulations shown in Fig.\ \ref{SPUR_SUBDIF}, in Fig.\ \ref{hFBM} the Hurst exponents were constant in time but random for each trajectory. They were assigned from the exponential distribution $p(H)=3.72 \exp(-3.72 H)$ which corresponds to the PDF of anomalous exponents $\mbox{PDF}(\alpha)=1.86 \exp(-1.86 \alpha)$ observed for experimental endosomal trajectories (see Fig.\ \ref{LOC_ALPHA_D}a). Hurst exponents are bounded, $H \in [0,1]$. The generalized diffusion coefficients were assigned from the power law distribution $f(D)=\gamma D_{min}^{\gamma} D^{-1-\gamma}$ with $\gamma=0.5$ as it was observed for endosomal trajectories (see Fig.\ \ref{LOC_ALPHA_D}b). The PDF of $D$ was truncated at $D_{min}=0.1$ and $D_{max}=100$. 

\subsection{Simulation of spatio-temporal heterogeneous FBM trajectories (Fig.\ \ref{SThFBM})}

In contrast to the spatially heterogeneous FBM trajectories, the anomalous exponent, $H$, and the generalized diffusion coefficient, $D$, of a single spatio-temporal heterogeneous FBM trajectory are not constant in time. $H$ randomly switches between $H=0.2$ and $H=0.6$ with switching rates $4/$s and $1/$s. At $t=0$, we choose $H=0.2$ or $H=0.6$ with equal probability. The diffusivities $D$ of the persistent regime $H=0.6$ were assigned from a power law distribution $f(D)=\gamma D_{min}^{\gamma} D^{-1-\gamma}$ with $\gamma=0.5$ as it was observed for experimental endosomal trajectories (see Fig.\ \ref{EMSD}). The PDF of $D$ was truncated at $D_{min}=0.0004$ and $D_{max}=20$. The diffusivities of the anti-persistent regime ($H=0.2$) were chosen as $0.2 D$. In order to reproduce the apparent subdiffusive behaviour of the experimental MSDs at longer times (see the discussion in the Results and Fig.\ \ref{SPUR_SUBDIF}), we couple the duration of trajectories with their diffusivities. We chose $T = 13 D^{-0.6}$ s which was truncated by $T_{min}=0.04$ s and $T_{max}=40$ s.

\section{Discussion}

It has been known for a long time that complex endosomal movement in eukaryotic cells consists of random switching from fast directed movement powered by molecular motors interspersed with periods of slow movement characterized by anomalous diffusion, rather than Brownian motion \cite{Rodriguez}. From our results, we find that standard anomalous diffusion with a single constant anomalous exponent and a single generalized diffusion coefficient for the ensemble cannot adequately describe the transport of endosomes. Instead, a time-space heterogeneous FBM with an anomalous exponent that switches between super-diffusive motion $\alpha>1$ and sub-diffusive motility $\alpha<1$ can do so. We conclude this because the local dynamics of endosomes display a whole spectrum of values (Figs.\ \ref{LOC_ALPHA_D}, \ref{H}). The anomalous exponents were found to follow an exponential probability distribution while the generalized diffusion coefficients are described by a power-law probability distribution. This result is robust since it does not depend on the method used to obtain these distributions. Both standard statistical analysis and the newly developed neural network \cite{ELife} trained to detect FBM corroborate these results. 

Another characteristic of the heterogeneous diffusion is the non-Gaussian distributions of displacements and displacement increments. Often, Laplace-like exponentially decaying densities were observed \cite{Metzler2020}, which were accompanied by apparent Brownian diffusion and exponential probability distributions of diffusion coefficients. In our data, we observed anomalous diffusion and power-law probability distributions of displacements and generalized diffusion coefficients respectively for experimental trajectories of endosomes. Theoretically, the distribution of displacements of spatially heterogeneous ensemble of FBM trajectories \cite{Chechkin} with the distribution of diffusion coefficient of a power law form $p\left(D_\alpha\right)\sim D_\alpha^{-1-\gamma}$, have power law tails of the form $P\left(X\right)\simeq\left|X\right|^{-1-2\gamma}$ (see Supplementary Note and Ref. \cite{Chechkin}). For experimental endosomal trajectories we find the power law distribution of local generalized diffusion coefficients and estimated $\gamma\sim0.5\ \pm0.1$ (Fig. \ref{LOC_ALPHA_D}b). This gives $1+2\ \gamma\sim2$ for the exponent of the power-law tail of power-law distributions which agrees well with the exponents found in the experimental PDFs (Figs.\ \ref{PDFX}, \ref{PDFDX}).

What are the factors promoting the appearance of the power-law distributions of the generalized diffusion coefficients? There are 3 possible reasons. (i) Power-law distribution of diffusion coefficients may indicate that endosomal networks exhibit scale free properties \cite{Julicher}, many small endosomes, and few large endosomes. Indeed, it was found in single-molecule experiments within the cell that fluctuations of molecular size generate a heterogeneous diffusion process with non-Gaussian distributions \cite{Lampo,SadoonWang,He}. Hence differences in their diameters generate distinct diffusive properties. (ii) Power-law distribution of local diffusivities of endosomes could be related to a power-law probability distribution of the empirical velocities of motor proteins which drive endosomes. This power law distribution of velocities originates from random attachments and detachments of motors to microtubules \cite{Hughes}. (iii) The power law distribution of $D_{\alpha_i^L}$ could be the non-specific interactions of endosomes with the endoplasmic reticulum and other organelles. Indeed, non-specific interactions were recently shown to generate heterogeneous diffusion of nanosized objects in mammalian cells \cite{Etoc}. It is possible that each or the interplay of all scenarios contribute to the observed power-law probability distributions of local generalized diffusion coefficients of endosomes. However, further experiments are required to verify this.

Can other anomalous diffusion models describe endosomal transport? There are three popular anomalous diffusion models for intracellular transport: the continuous time random walk, the generalized Langevin equation and Fractional Brownian motion. The equivalence of the ensemble and time averaged MSDs (averaged over all trajectories) shows that the endosomal transport is ergodic which rules out the continuous time random walk model as it is non-ergodic. The remaining FBM and generalized Langevin equation models are both ergodic but differ by their physical assumptions. The generalized Langevin equation model assumes local equilibrium while the FBM model describes out of equilibrium motion. Since molecular motors drive dynamics of endosomes out of equilibrium, we can also rule out the generalized Langevin dynamics. This suggests that the spatio-temporal heterogeneous FBM is the most appropriate anomalous diffusion model to describe endosomal dynamics. The ensemble of spatio-temporal hFBM trajectories display ergodicity e.g., the equivalence of the ensemble and ensemble-time averaged MSDs (TMSDs averaged over all trajectories).

In conclusion, for the first time we have shown that the anomalous intracellular transport of endosomes is described the spatio-temporal heterogeneous ensemble of FBM motions. We find that endosomal displacements and increments both have strongly non-Gaussian power laws distributions. Analysing local endosomal dynamics, we find that it is described by a spectrum of exponentially distributed anomalous exponents and the power-law distributed generalized diffusion coefficients. Such heterogeneity of endosomal transport has implications for sorting and delivering molecules for different biochemical reactions in different locations inside living cells. Heterogeneous dynamics of endosomes in space and time has a huge impact on these diffusion-limited reactions: it broadens the distribution of the first-passage time to a reaction event and increases the likelihood of both short and long trajectories hitting targets \cite{Rodriguez,Grebenkov2018,Godec}. Both effects haves a big impact on biochemical reaction rates, for example signalling of Rab proteins on endosomes. The dynamic properties of other intracellular organelles, such as mitochondria and lipid droplets, share many similarities with endosomal transport \cite{PLOS,Fedotov,Kenwright,Detmer} and we anticipate their transport properties will also be heterogeneous.

\begin{acknowledgments}
The authors thank: NK and SF acknowledge financial support from EPSRC Grant No. EP/V008641/1. DH acknowledges financial support from the Wellcome Trust Grant No. 215189/Z/19/Z. GP is supported by the Basque Government through the BERC 2018-2021 programs and by the Spanish Ministry of Economy and Competitiveness MINECO through the BCAM Severo Ochoa excellence accreditation SEV-2017-0718.
\end{acknowledgments}

\appendix

\setcounter{figure}{0}
\makeatletter 
\renewcommand{\thefigure}{S\@arabic\c@figure}
\makeatother

\begin{figure}[ht]
\centering
\includegraphics[scale=0.3]{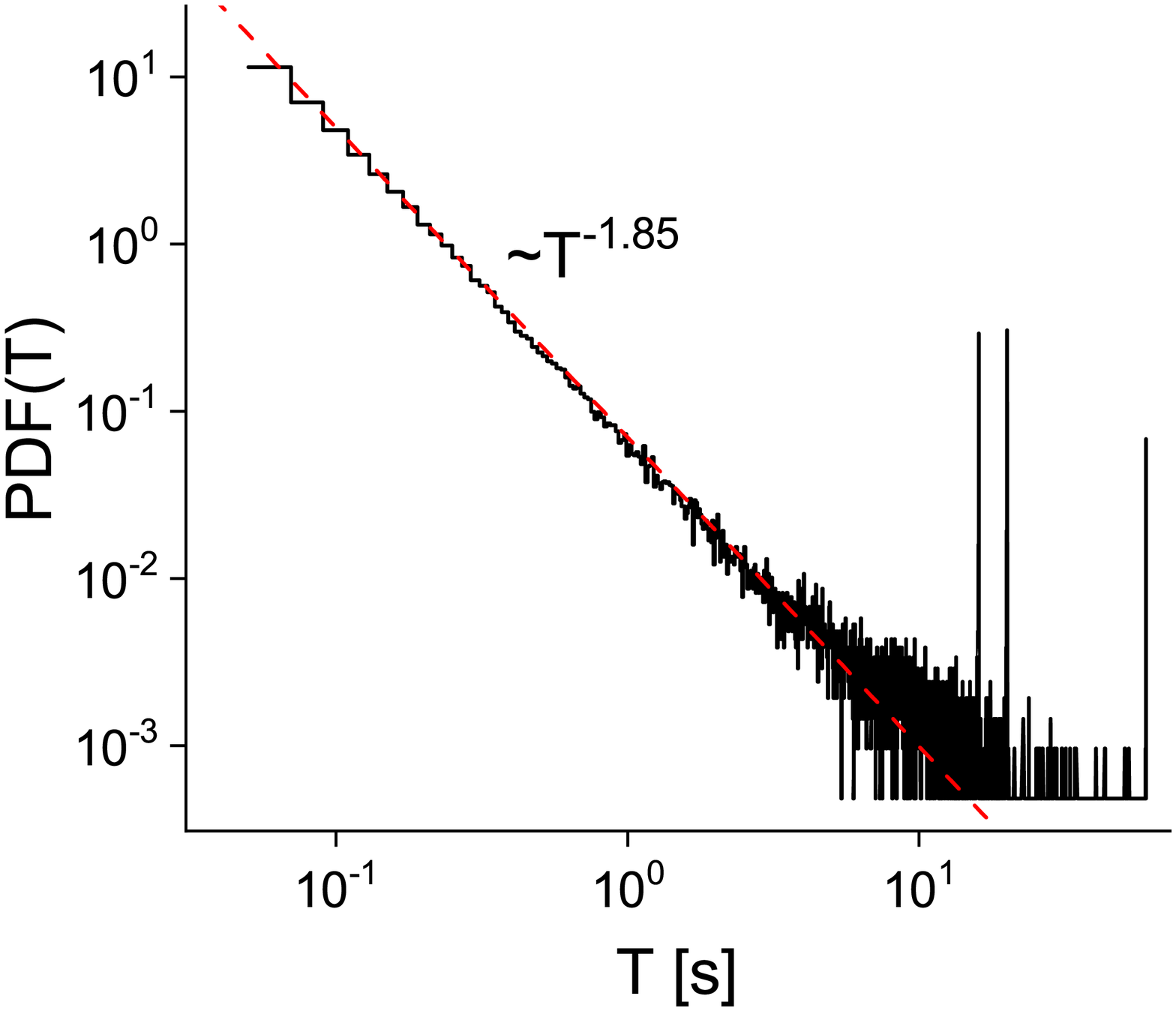}
\caption{Probability density function of the duration of endosomal trajectories, $T$. The dashed line represents the power law fit, $PDF(T) \sim T^{-1.85}$. 
}
\label{PDFT}
\end{figure}
\begin{figure}[ht]
\centering
\includegraphics[scale=0.3]{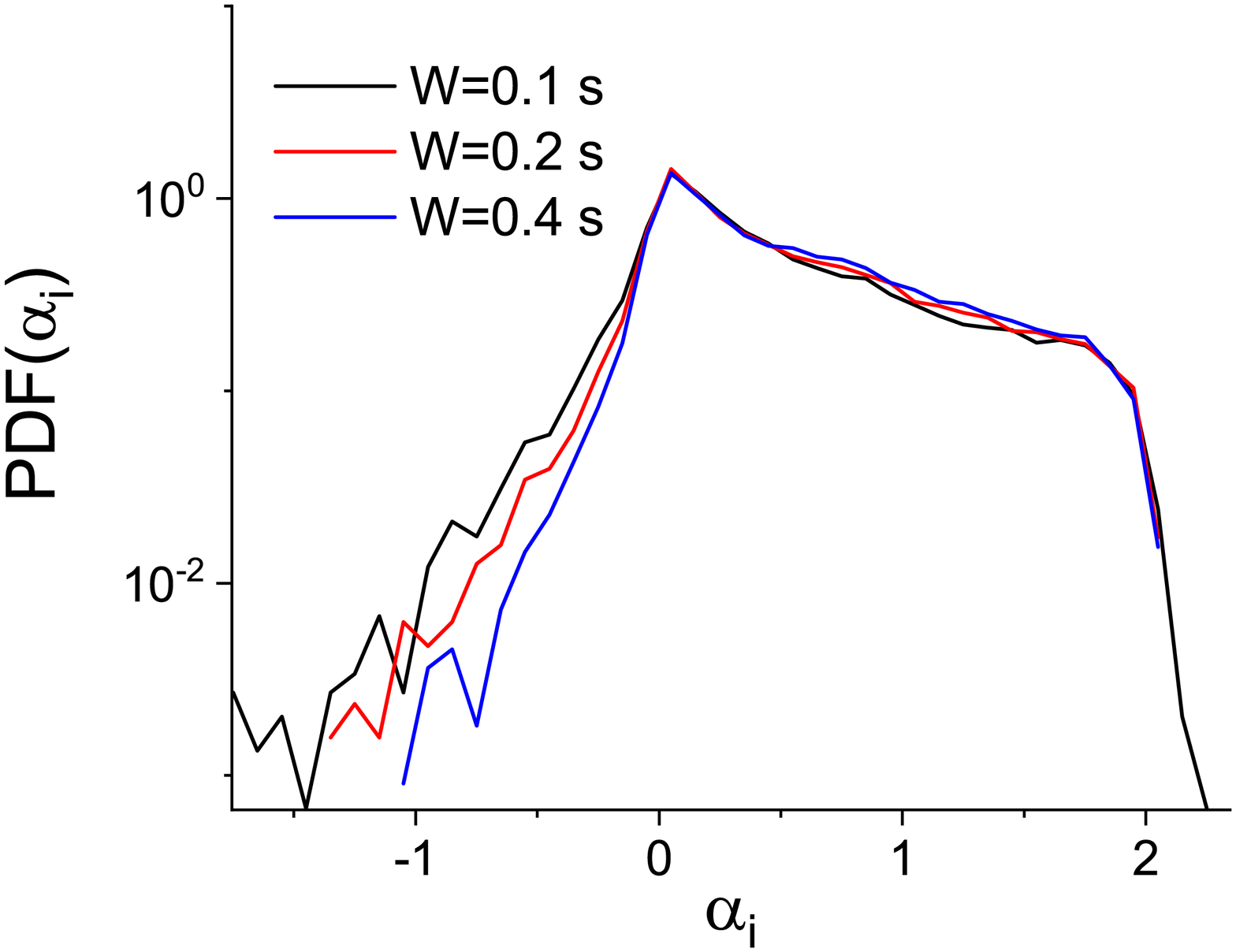}
\caption{Probability density function of anomalous exponents, $\alpha_i$, extracted from experimental endosomal trajectories does not depend on window size $W$ in the region $\alpha_i \in (0,2)$. Different curves correspond to different window size. PDFs were obtained at $t=0.1$ s for different window size $W$ indicated in the legend. Negative values of $\alpha_i$ and $\alpha_i>2$ which are artefacts of fitting of TMSDs gradually diminish with increased window size $W$.
}
\label{PDFT}
\end{figure}

\begin{figure*}[ht]
\centering
\includegraphics[scale=0.22]{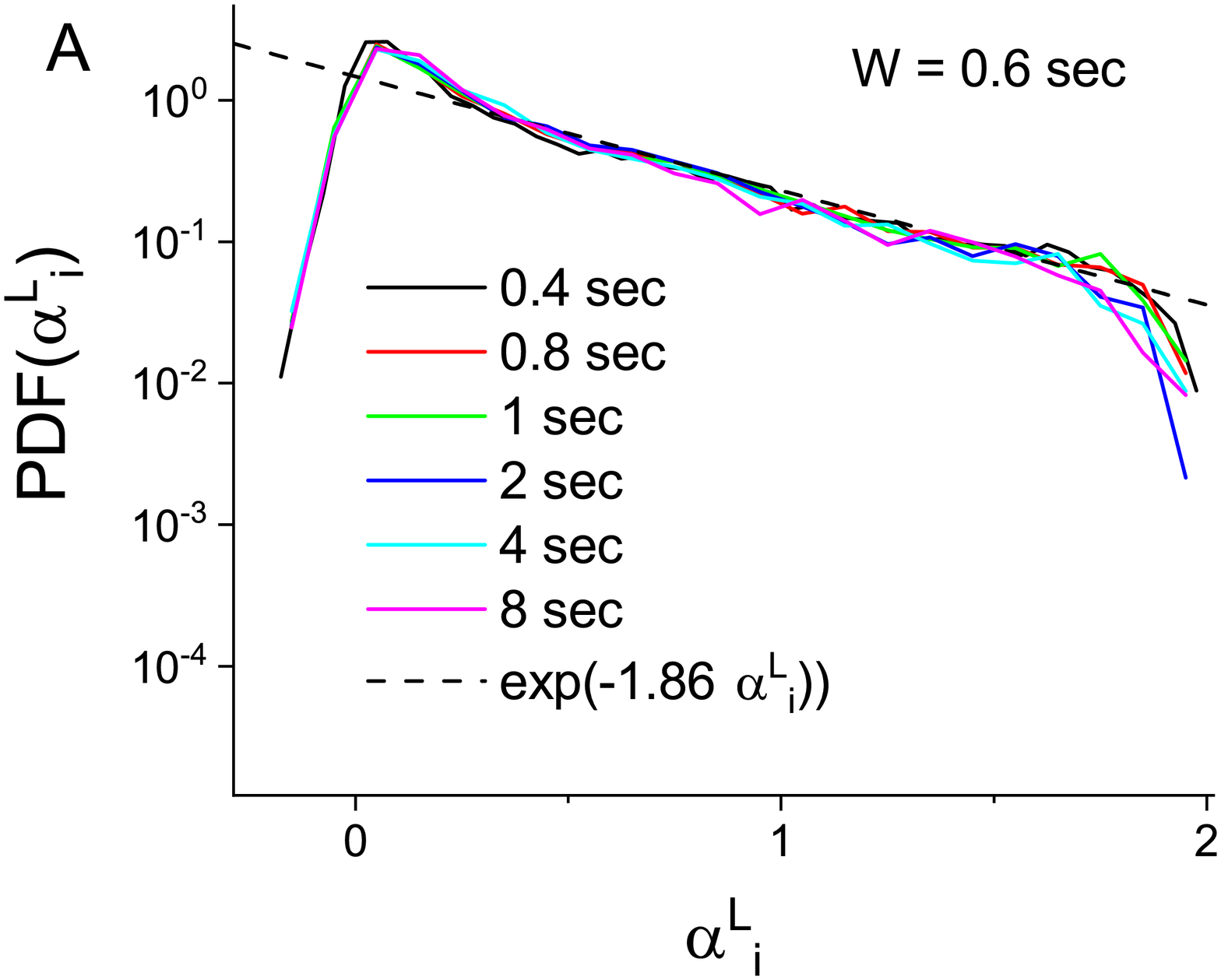}
\includegraphics[scale=0.22]{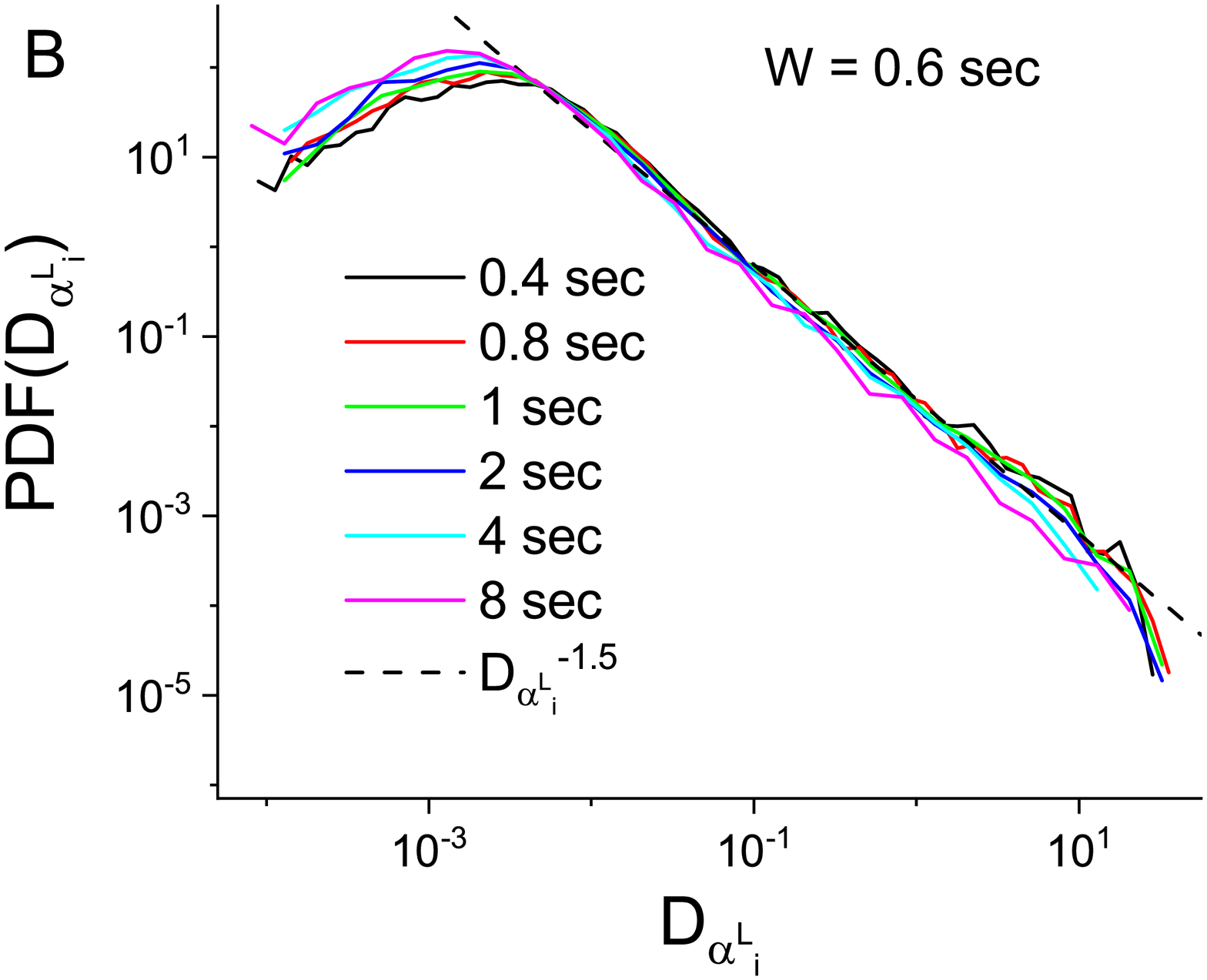}
\includegraphics[scale=0.22]{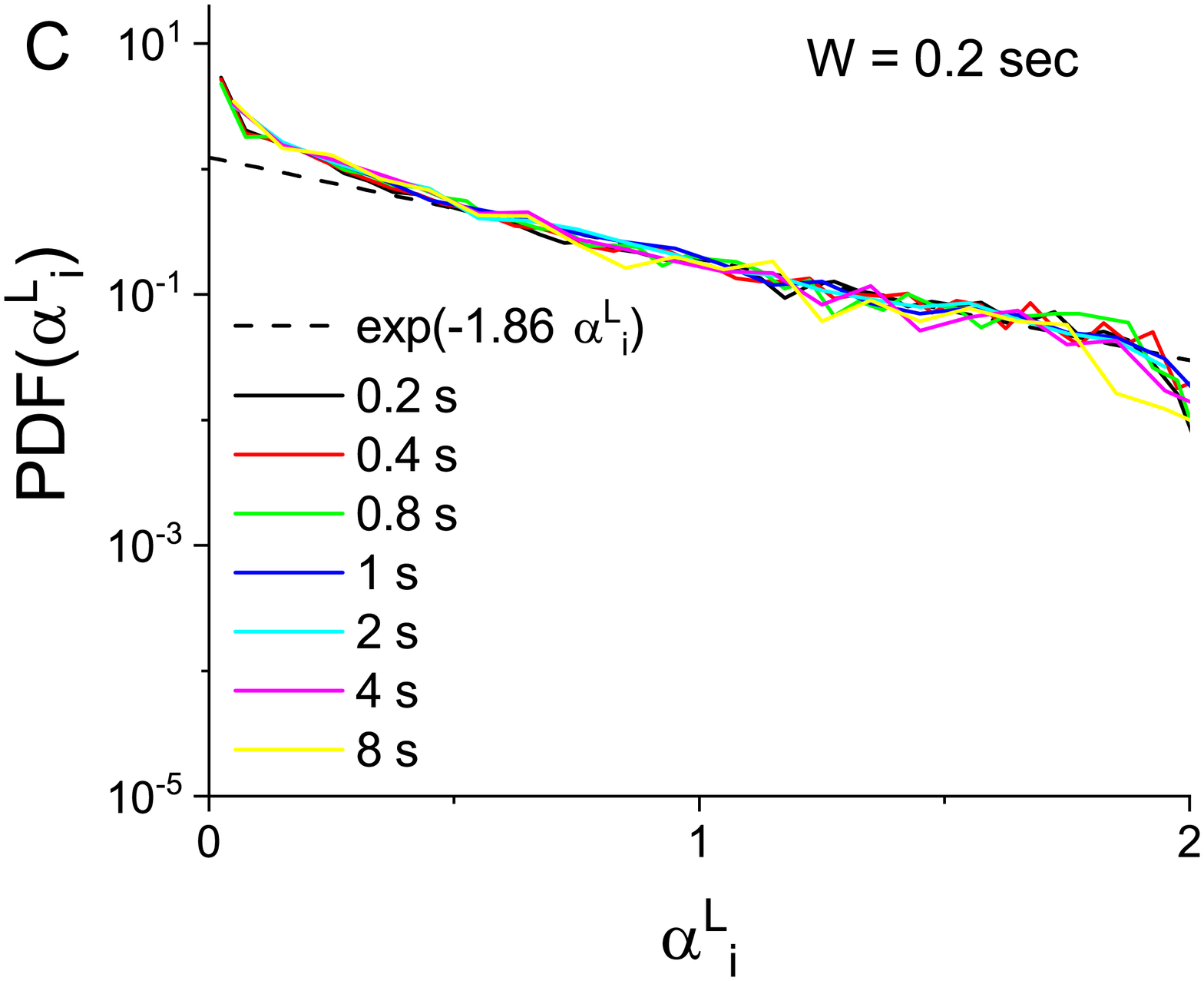}
\includegraphics[scale=0.22]{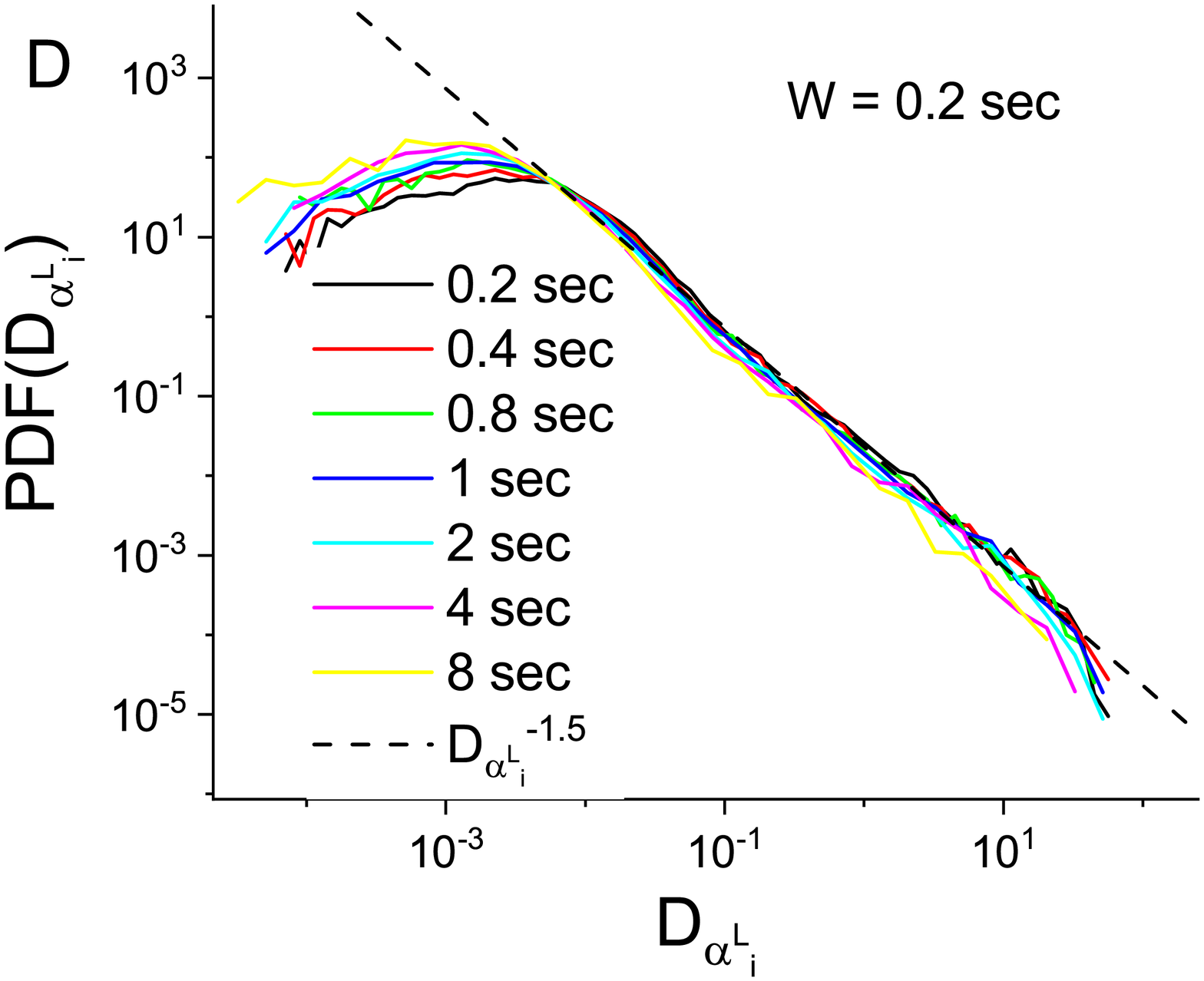}
\caption{Probability distributions of local anomalous exponents $\alpha^{L}_i$ and local generalized diffusion coefficient $D_{\alpha^{L}_i}$ estimated from local TMSDs (LTMSDs) are robust with respect to changing window size $W$. Figures (A), (B) show PDFs for window size $W=0.6$ s and (C), (D) for window size $W=0.2$ s.  
}
\label{LOC_ALPHA_D_SET2_SET3}
\end{figure*}

\begin{figure*}[ht]
\centering
\includegraphics[scale=0.3]{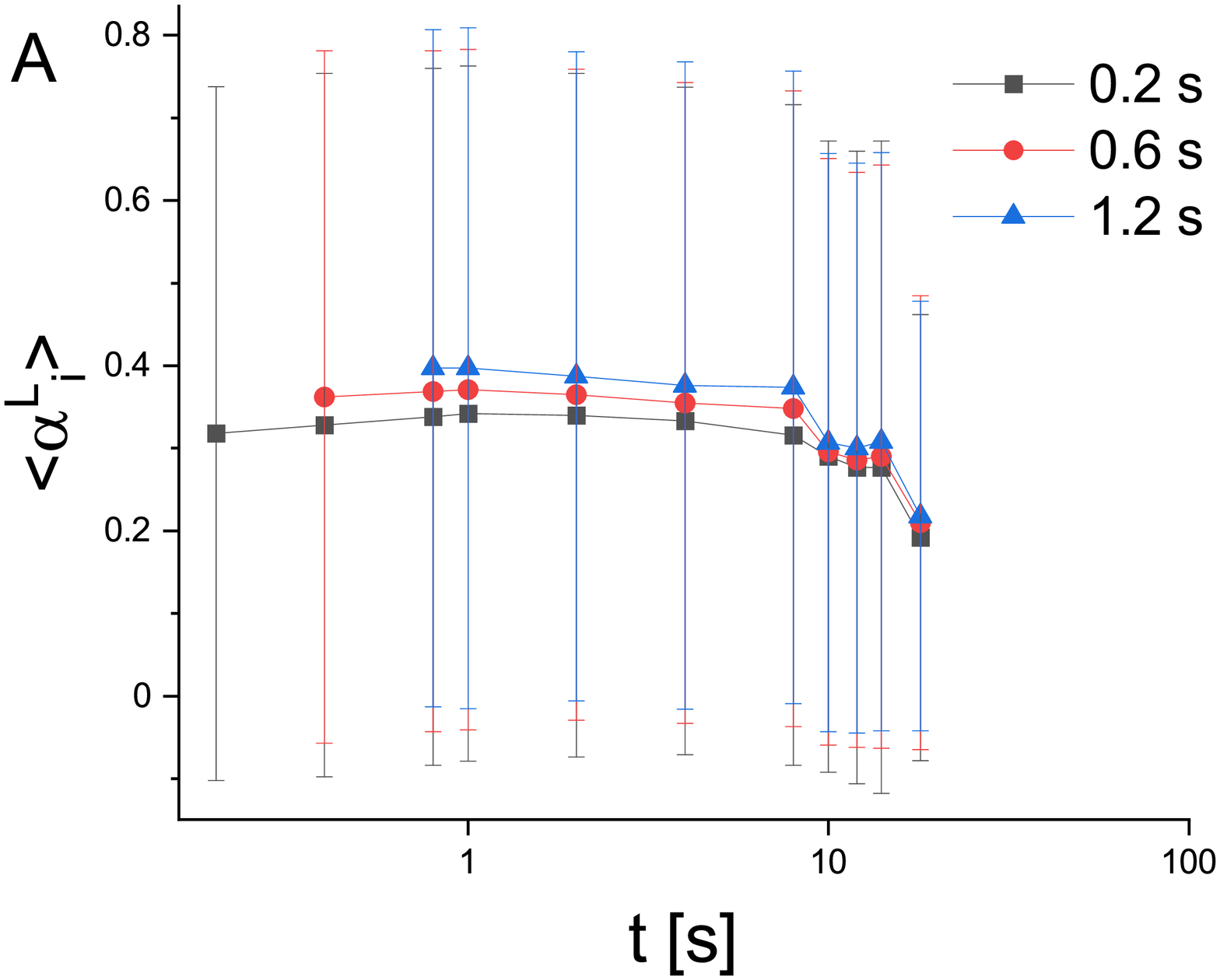}
\includegraphics[scale=0.3]{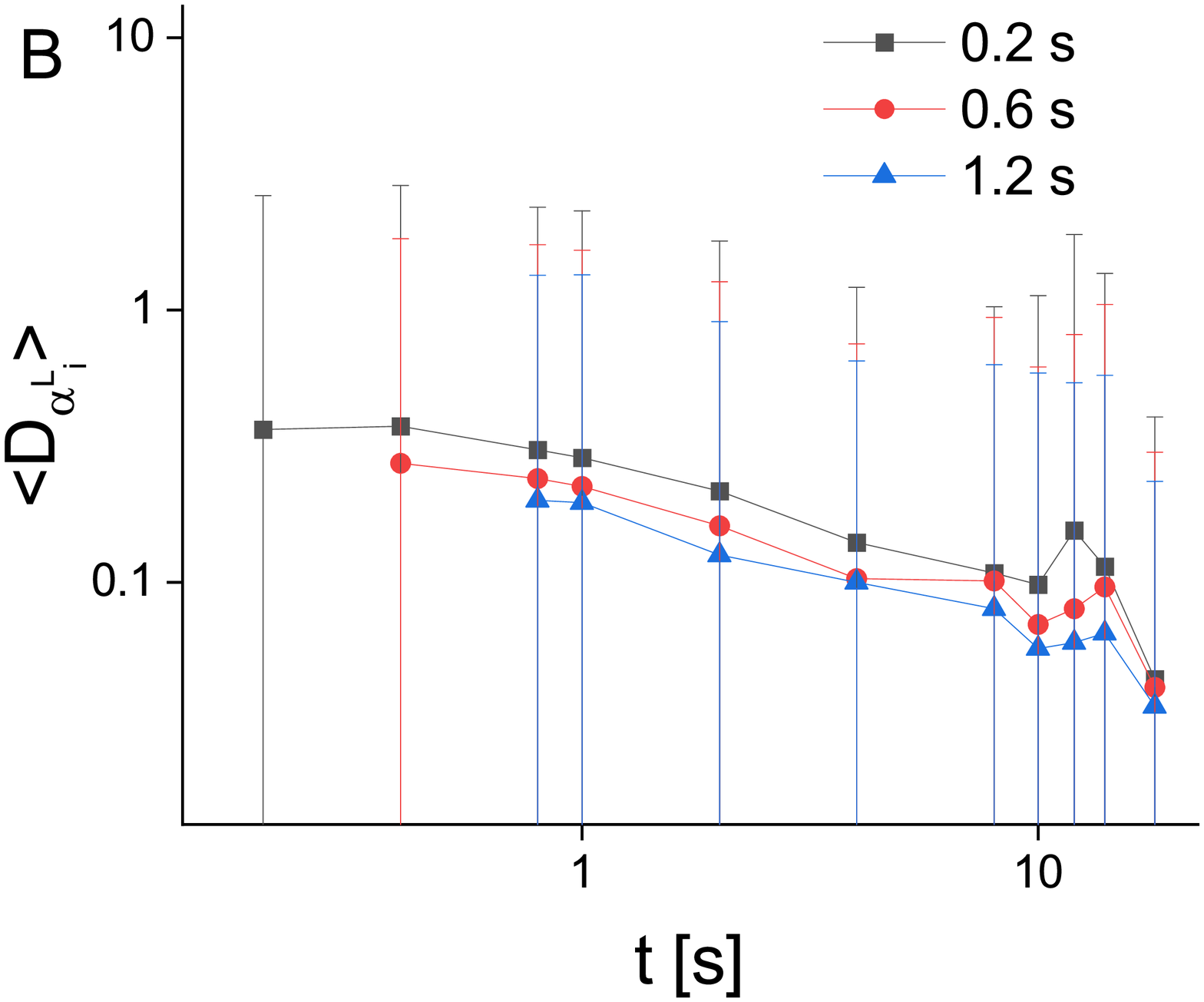}
\caption{The average local anomalous exponent $\left< \alpha^{L}_i \right>$ (A) stays constant for $t<8$ s while the average local generalized diffusion coefficient $\left< D_{\alpha^{L}_i} \right>$ (B) decrease with time. For $t>8$ s both $\left< \alpha^{L}_i \right>$ and $\left< D_{\alpha^{L}_i} \right>$ decrease with time. Different curves correspond to window size $0.2$ s, $0.6$ s and $1.2$ s. The error bars in both panels represent standard deviation.
}
\label{AV_ALF_D}
\end{figure*}

\begin{figure*}[ht]
\centering
\includegraphics[
  width=14cm,
  keepaspectratio,
]{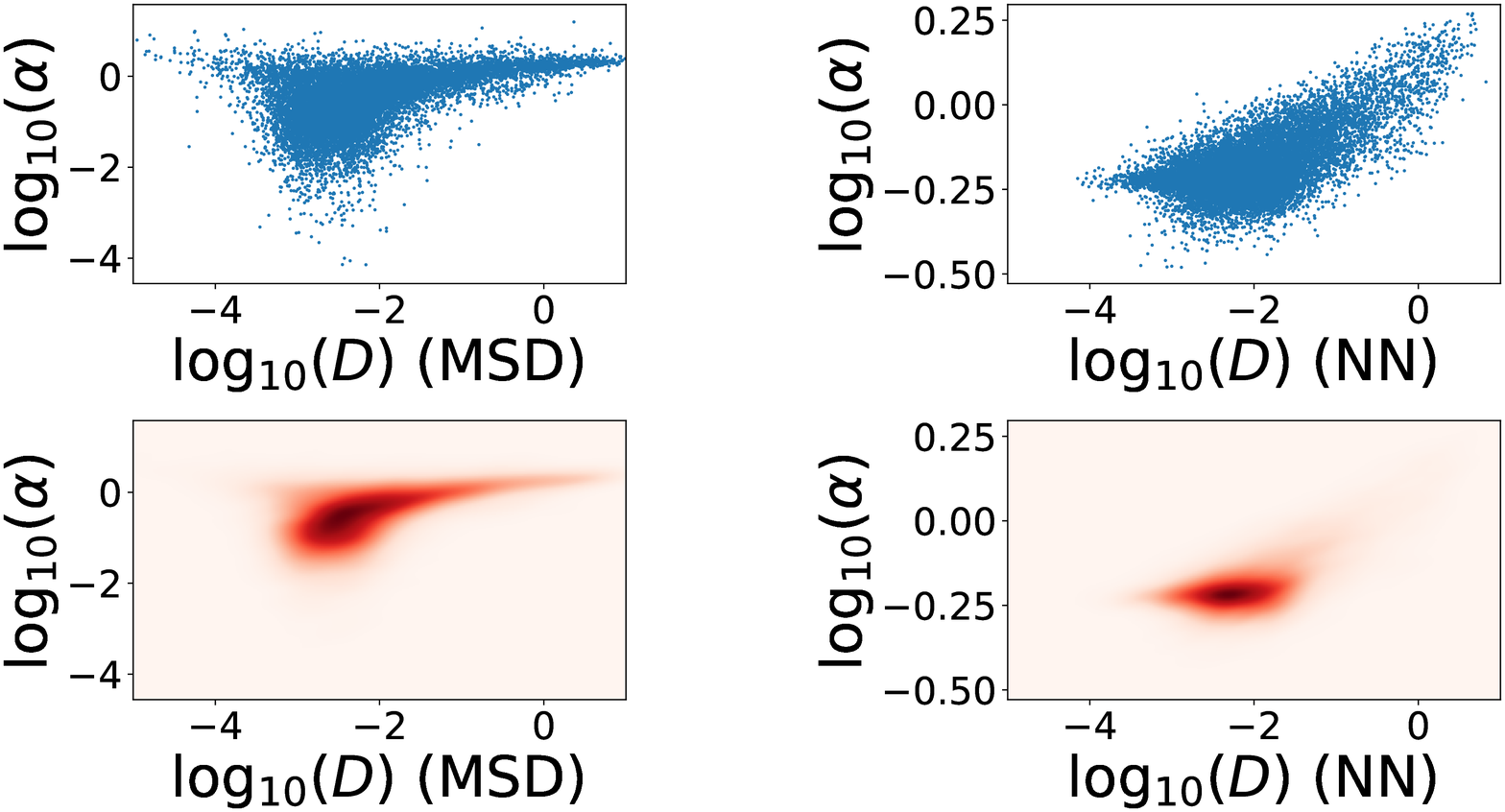}
\caption{Anomalous exponents $\alpha$ and generalized diffusion coefficients $D$ are positively correlated. (Top row) Scatter plots of $\alpha$ against $D$ calculated from each trajectory using two different methods: power-law fitting to MSDs; and neural network estimation. (Bottom row) Kernel density estimation heat maps for the scatter plots in the top row.}
\label{D_ALPHA_COR}
\end{figure*}

\begin{figure*}[ht]
\centering
\includegraphics[scale=0.3]{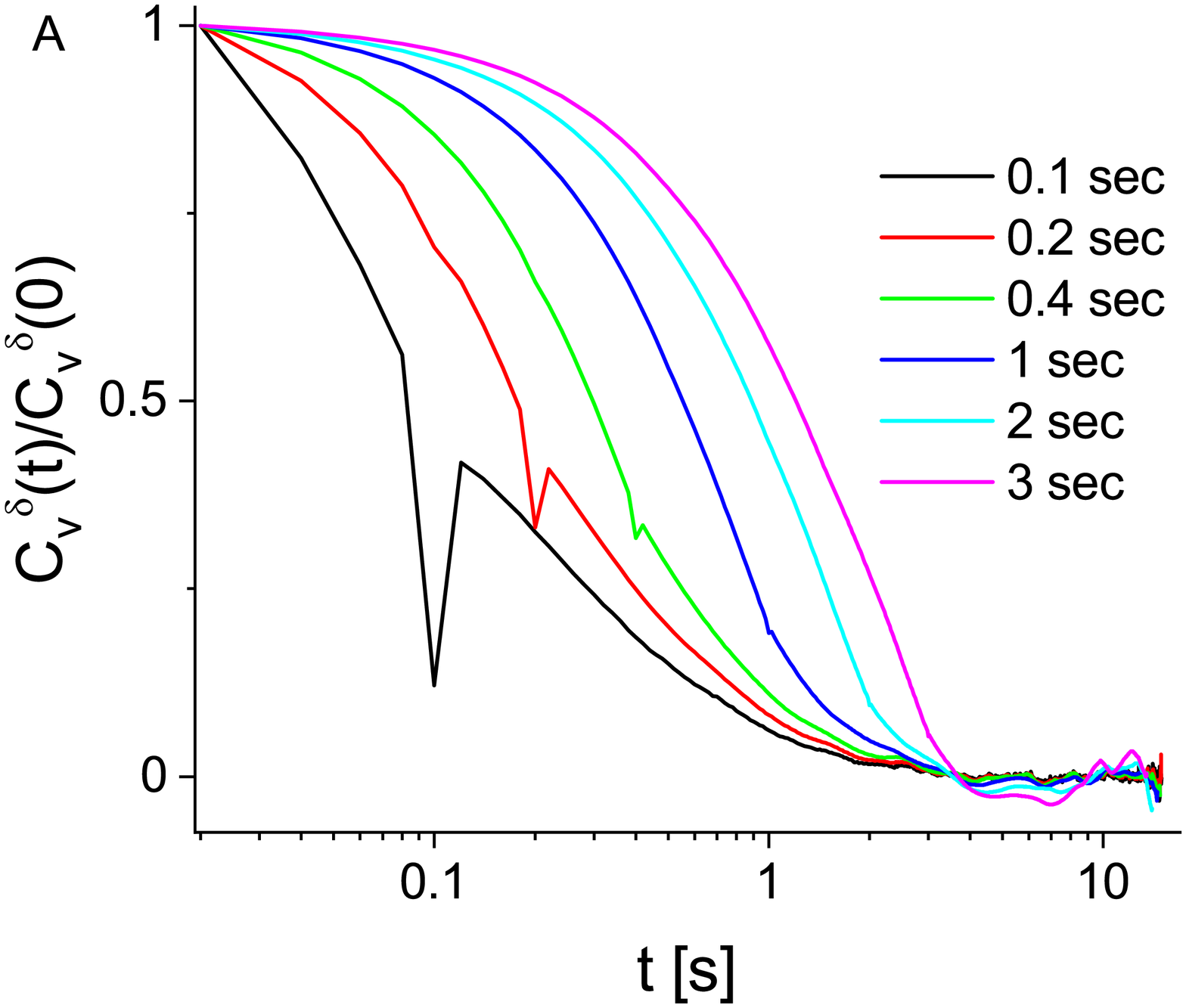}
\includegraphics[scale=0.3]{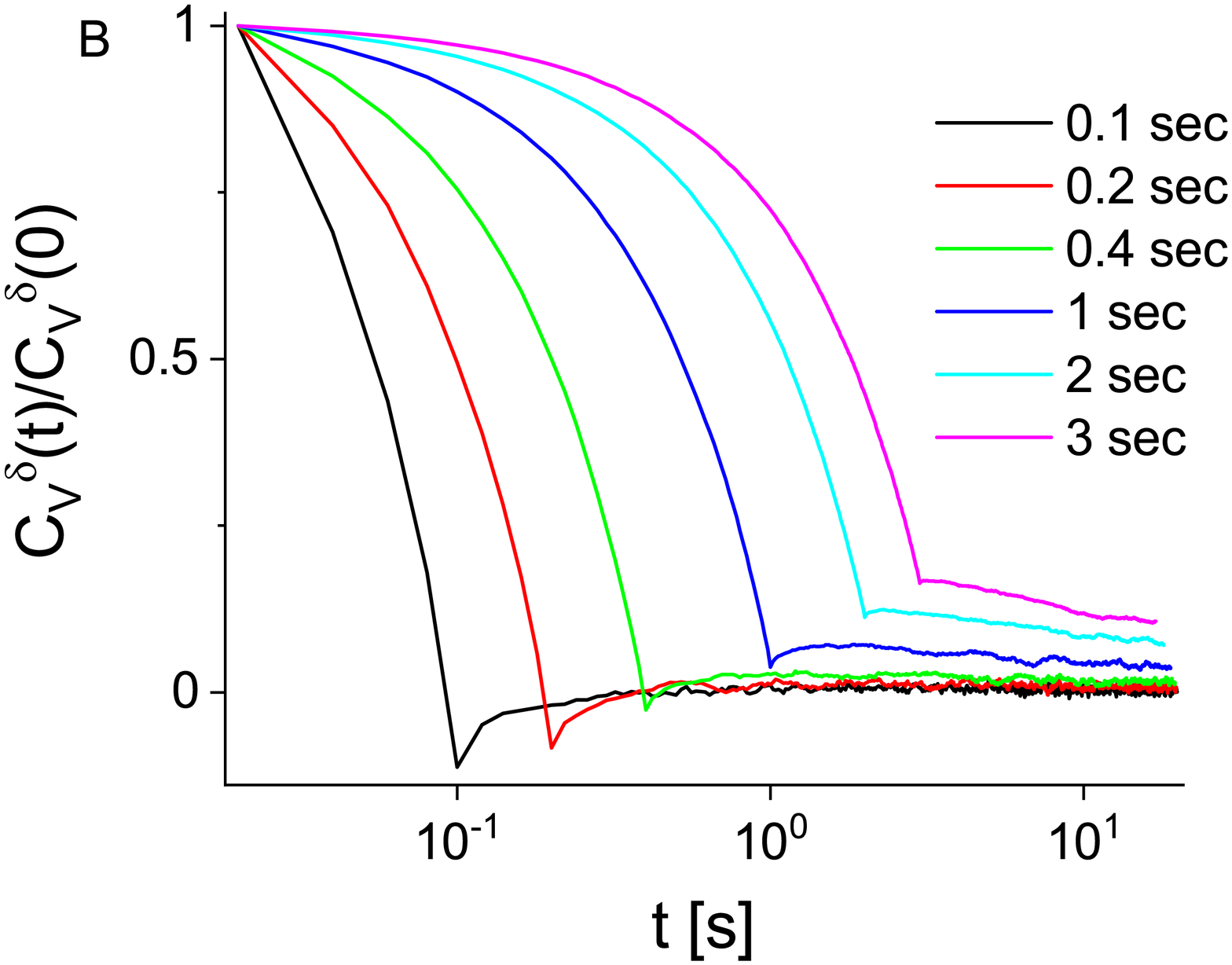}
\caption{Normalized ensemble averaged velocity auto-correlation function (EVACF) of experimental trajectories (A) and heterogeneous ensemble of noiseless FBM trajectories with exponential distribution of Hurst exponents and power law distribution of generalized diffusion coefficients (B). Values of $\delta$ are given in the legend. A common behaviour is observed: negative peaks at $t=\delta$ gradually disappearing with time. The FBM trajectories were generated without measurement noise.}
\label{VACF}
\end{figure*}

\begin{figure*}[ht]
\centering
\includegraphics[scale=0.3]{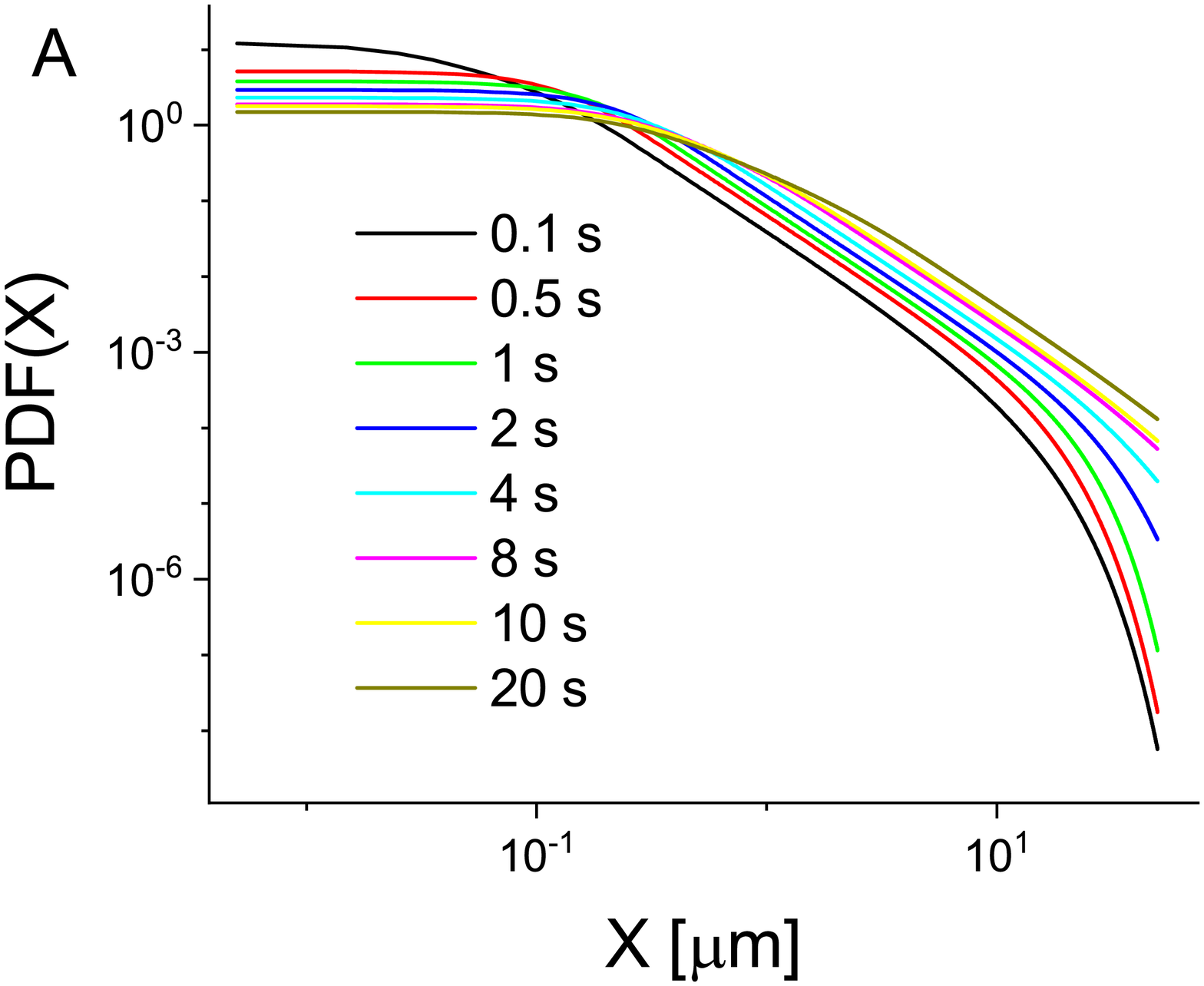}
\includegraphics[scale=0.3]{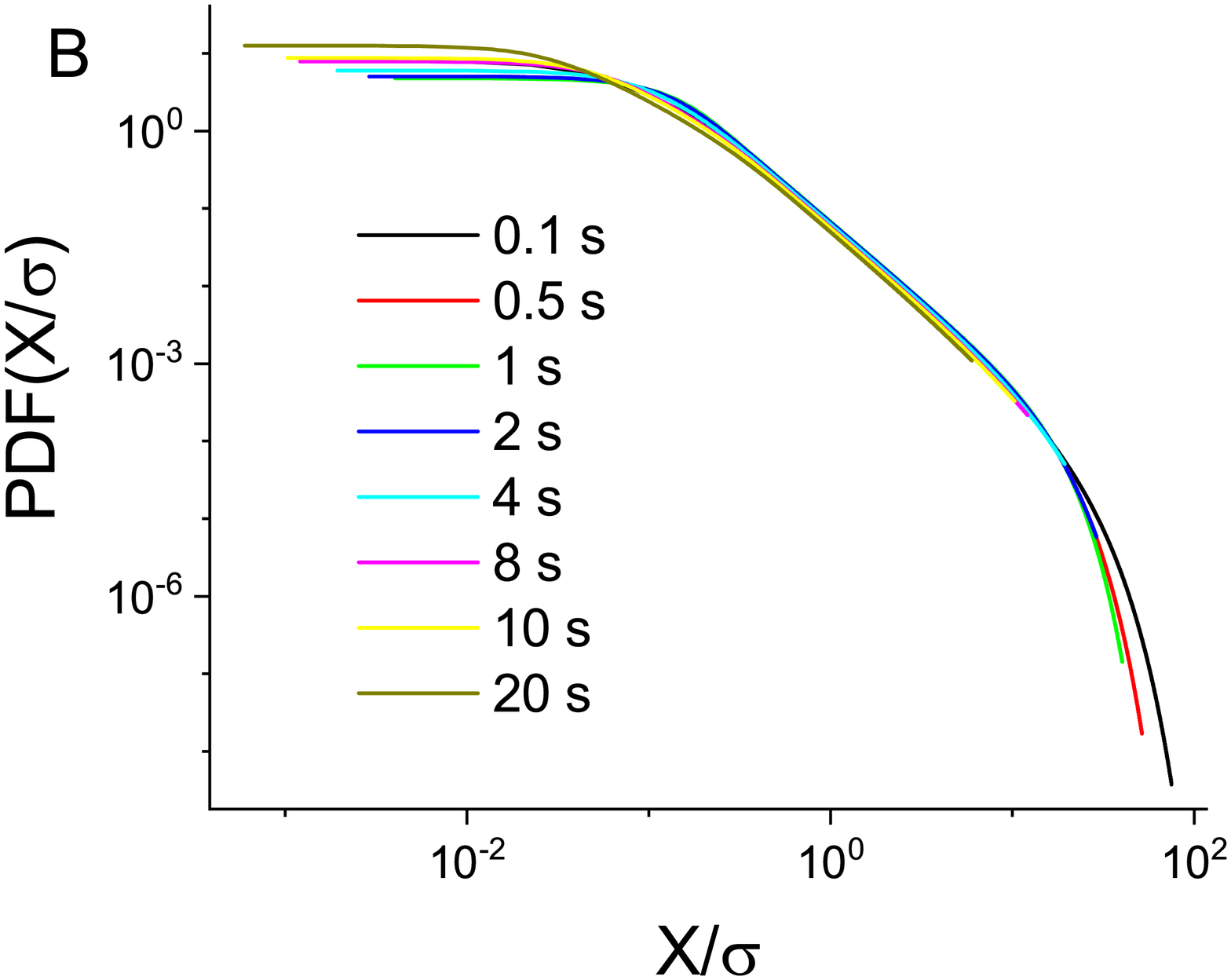}
\caption{Distributions of displacements (A) and distributions of displacements scaled by their standard deviation (B) for the superstatistical ensemble of FBM trajectories with the exponentially distributed anomalous exponents $\alpha$ and power-law distributed generalized diffusion coefficients $D$. PDFs were obtained by numerical integration of Eq.\ (\ref{P_2}) at different times indicated in the legends. The distributions of displacements scaled by their standard deviation in (B) show collapse in the central part.}
\label{SFBM}
\end{figure*}

\section*{Appendix A}

\setcounter{equation}{0}

\subsection{Derivation of the power-law distribution of displacements and increments for heterogeneous ensemble of FBM}

For simplicity we consider the 1D case which is easily generalized to 2D. For the FBM with a constant generalized diffusion coefficient $D$ and anomalous exponent $\alpha$, the distribution of displacements is Gaussian \cite{Metzler2014}: 
\begin{equation}
G(x,t | D,\alpha) = \frac{1}{\sqrt{4 \pi D t^{\alpha}}} \exp 
\left( - \frac{x^2}{4 D t^{\alpha}} \right) .
\label{G}
\end{equation}
In a superstatistical ensemble, each trajectory is assumed to explore a local spatial patch which is characterized by its own constant diffusivity $D$ and anomalous exponent $\alpha$. The values of $D$ are drawn from the PDF $f(D)$ and values of $\alpha$ from the PDF $g(\alpha)$. The resulting distribution of displacements of the heterogeneous ensemble of FBM trajectories is given by the convolution:
\begin{equation}
P(x,t) = \int_{0}^{2} \int_{0}^{\infty} G(x,t | D,\alpha) f(D) g(\alpha) dD d\alpha.
\label{P}
\end{equation}
We consider the PDFs of $\alpha$ and $D$ which were suggested by the experimental data. Specifically, we choose the power law PDF of $D$: 
\begin{equation}
p(D) = \frac{\gamma D^{-1-\gamma}}{D_{-}^{-\gamma} - D_{+}^{-\gamma}}, 
\label{PDF_D}
\end{equation}
and the exponential PDF of $\alpha$: 
\begin{equation}
h(\alpha) = \frac{e^{-\alpha/\left<\alpha\right>}}{\left<\alpha\right> (1 - e^{-2/\left<\alpha\right>})}. 
\label{PDF_ALF}
\end{equation}
From experimental trajectories we estimated $\gamma \sim 0.5$ and $\left<\alpha\right> \sim 0.53$ (Fig.\ 5). The distribution of displacements Eq.\ \ref{P} then reads:
\begin{widetext}
\begin{equation}
P(x,t) = \frac{\gamma 4^{\gamma}}{\sqrt{\pi} (D_{-}^{-\gamma} - D_{+}^{-\gamma}) (1 - e^{-2/\left<\alpha\right>})} \int_{0}^{2} e^{-\alpha \left(\frac{1}{\left<\alpha\right>} - \frac{\gamma \ln{t}}{2}\right)} \left[ \Gamma \left( \gamma + \frac{1}{2}, \frac{x^2}{4 D_{+} t^{\gamma}}\right) - \Gamma \left( \gamma + \frac{1}{2}, \frac{x^2}{4 D_{-} t^{\gamma}}\right)  \right] d\alpha.
\label{P_2}
\end{equation}
\end{widetext}
Fig.\ \ref{SFBM} (A) illustrates PDFs of $x$ at different time which were obtained by numerical integration of Eq. \ref{P_2}. The PDFs of the scaling variable $X/\sigma$ in Fig.\ \ref{SFBM} (B) show the data collapse in the central part.

Considering the Fourier-Laplace transforms and expanding integrals in Eq.\ (\ref{P}) at long time and space regime, the power law behaviour of the distribution of displacements is \cite{Wang}:
\begin{equation}
P(x,t) \simeq |x|^{-1-2 \gamma}.
\label{P1}
\end{equation}
Since for the distribution of increments for standard FBM is also Gaussian \cite{Metzler2014}, a similar derivation holds for the power law distribution of increments for the superstatistical ensemble of FBM trajectories.

\subsection{Derivation of the ensemble averaged mean squared displacement (EMSD) for superstatistical ensemble of FBM}

The EMSD is defined as:
\begin{equation}
\left< x^{2}(t) \right> = \int_{-\infty}^{\infty} x^{2} P(x,t) dx. 
\label{X2}
\end{equation}
Using the expression for $P(x,t)$ and PDFs $f(D)$, $h(\alpha)$, we obtain the following expression:
\begin{equation}
\left< x^{2}(t) \right> = \frac{2 \left<D\right>}{1-\left<\alpha\right>\ln{t}} \frac{1 - e^{-2/\left<\alpha\right>} t^2}{1 - e^{-2/\left<\alpha\right>}},
\label{X2_2}
\end{equation}
where 
\begin{equation}
\left<D\right> = \frac{\gamma}{1-\gamma} \frac{(D_{+}^{1-\gamma} - D_{-}^{1-\gamma})}{(D_{-}^{-\gamma} - D_{+}^{-\gamma})}.
\end{equation}

\subsection{ The ensemble and the time averaged velocity auto-correlation functions for superstatistical ensemble of FBM}

The velocity auto-correlation function (ACF) averaged over an ensemble of trajectories (EACF) is calculated as:
\begin{equation}
C_{v}^{\delta}(t) = \left<\vec{v}(t) \vec{v}(0) \right> .
\label{evacf}
\end{equation}
where $\vec{v}=\frac{\vec{r}(t+\delta)-\vec{r}(t)}{\delta}$.
The time averaged ACF along a single trajectory (TACF) is defined as: 
\begin{equation}
\overline{C_{v}^{\delta}(t)} = \frac{ \int_{0}^{T-t - \delta} \vec{v}(t'+t) \vec{v}(t') dt'}{T-t - \delta}.
\label{tvacf}
\end{equation}

\section*{Appendix B: Supplementary figures}

Figure S1: the probability density function of the duration of experimental endosomal trajectories, $T$, is shown. The power law, $PDF(T) \sim T^{-1.85}$, fits well the data. 

Figure S2: The PDFs of anomalous exponents $\alpha_i$ extracted from experimental endosomal trajectories are robust. The Figure shows PDFs obtained at $t=0.1$ s for different window size $W$ indicated in the legend. In the region $\alpha_i \in (0,2)$, PDFs stay the same. The negative values of $\alpha_i$, which are artefacts of fitting of TMSDs, gradually diminish with increased window size $W$.

Figure S3: The PDFs of local anomalous exponents $\bar{\alpha}$ and local generalized diffusion coefficient $D_{\bar{\alpha}}$ estimated from local TMSDs (LTMSDs) of experimental endosomal trajectories are robust. PDFs of $\bar{\alpha}$ and $D_{\bar{\alpha}}$ obtained for different window size $W$ are the same.

Figure S4: The average over all trajectories local anomalous exponents $\left< \bar{\alpha} \right>$ (panel (A)) stay constant for $t<8$ s while the average local generalized diffusion coefficients $\left< D_{\bar{\alpha}} \right>$ (panel (B)) decrease with time. This suggests that longer trajectories have smaller generalized diffusion coefficients, see Fig. 4 and Fig. 8a in the main text. For $t>8$ s both $\left< \bar{\alpha} \right>$ and $\left< D_{\bar{\alpha}} \right>$ decrease with time which could be due to limited statistics.

Figure S5: Anomalous exponents $\alpha$ and generalized diffusion coefficients $D$ are positively correlated. (Top row) Scatter plots of $\alpha$ against $D$ calculated from each trajectory using two different methods: power-law fitting to MSDs; and neural network estimation. (Bottom row) Kernel density estimation heat maps for the scatter plots in the top row. 

Figure S6: Normalized ensemble averaged velocity auto-correlation function (EVACF) of experimental trajectories (A) and heterogeneous ensemble of FBM trajectories with exponential distribution of Hurst exponents and power law distribution of generalized diffusion coefficients (B). Values of $\delta$ are given in the legend. A common behaviour is observed: negative peaks at $t=\delta$ gradually disappearing with time. The FBM trajectories were generated without measurement noise. 

Figure S7: Distributions of displacements (A) and distributions of displacements scaled by their standard deviation (B) for the superstatistical ensemble of FBM trajectories with the exponentially distributed anomaous exponents $\alpha$ and power-law distributed generalized diffusion coefficients $D$. PDFs were obtained by numerical integration of Eq.\ (\ref{P_2}) at different times indicated in the legends. The distributions of displacements scaled by their standard deviation in (B) show collapse in the central part.


\end{document}